\documentclass[12pt]{article}
\usepackage{amsmath}
\usepackage{graphicx}
\usepackage{enumerate}
\usepackage{bm, bbm, mathtools ,psfrag, amsfonts, colonequals}
\usepackage[ruled,vlined,noend]{algorithm2e}
\SetKwInput{KwInput}{Input}
\usepackage{algorithmic}
\usepackage[usenames, dvipsnames]{color}
\usepackage[font={footnotesize}]{caption,subcaption}
\usepackage{hyperref}
\usepackage{natbib}
\mathtoolsset{showonlyrefs=true}
\usepackage{wrapfig}

\usepackage{xr}
\makeatletter
\newcommand*{\addFileDependency}[1]{
\typeout{(#1)}
\@addtofilelist{#1}
\IfFileExists{#1}{}{\typeout{No file #1.}}
}\makeatother

\newcommand*{\myexternaldocument}[1]{%
\externaldocument{#1}%
\addFileDependency{#1.tex}%
\addFileDependency{#1.aux}%
}
\myexternaldocument{JASA-supplement}

\newcommand{\blind}{1}

\newcommand{\bu}{\mathbf{u}}
\newcommand{\bl}{\mathbf{l}}
\newcommand{\ba}{\mathbf{a}}
\newcommand{\be}{\mathbf{e}}

\newcommand{\bb}{\mathbf{b}}
\newcommand{\bs}{\mathbf{s}}

\newcommand{\bx}{\mathbf{x}}
\newcommand{\by}{\mathbf{y}}

\newcommand{\br}{\mathbf{r}}
\newcommand{\bw}{\mathbf{w}}

\newcommand{\bz}{\mathbf{z}}

\newcommand{\bA}{\mathbf{A}}

\newcommand{\bG}{\mathbf{G}}

\newcommand{\bL}{\mathbf{L}}

\newcommand{\bI}{\mathbf{I}}
\newcommand{\bD}{\mathbf{D}}
\newcommand{\bH}{\mathbf{H}}

\newcommand{\bV}{\mathbf{V}}

\newcommand{\bfzero}{\mathbf{0}}

\newcommand{\bfgamma}{\bm{\gamma}}
\newcommand{\bfmu}{\bm{\mu}}

\newcommand{\bfSigma}{\bm{\Sigma}}

\newcommand{\bfPsi}{\bm{\Psi}}

\newcommand{\diag}{diag}

\newcommand{\order}{\mathcal{O}}
\newcommand{\normal}{\mathcal{N}}

\DeclareMathOperator*{\argmin}{arg\,min}
\DeclareMathOperator*{\argmax}{arg\,max}
\newcommand{\diff}{\mathrm{d}}

\newcommand{\defeq}{\vcentcolon=}

\usepackage{amsthm}
\newtheoremstyle{propstyle} 
    {2mm}                    
    {1mm}                    
    {\itshape}                   
    {}                           
    {\scshape}                   
    {.}                          
    {.5em}                       
    {}  
\theoremstyle{propstyle}

\theoremstyle{propstyle}

\theoremstyle{propstyle}
\newtheorem{proposition}{Proposition}
\theoremstyle{propstyle}

\theoremstyle{propstyle}

\theoremstyle{propstyle}

\theoremstyle{propstyle}

\begin{document}

\def\spacingset#1{\renewcommand{\baselinestretch}%
{#1}\small\normalsize} \spacingset{1}


\if1\blind
{
  \title{\bf Linear-Cost Vecchia Approximation of Multivariate Normal Probabilities}
  \author{Jian Cao \hspace{.2cm}\\
    Department of Mathematics, University of Houston \\ \texttt{jcao21@uh.edu}\\
    and \\
    Matthias Katzfuss \\
    Department of Statistics, University of Wisconsin--Madison \\ \texttt{katzfuss@gmail.com}}
  \maketitle
} \fi

\if0\blind
{
  \bigskip
  \bigskip
  \bigskip
  \begin{center}
    {\LARGE\bf Linear-Cost Vecchia Approximation of Multivariate Normal Probabilities}
\end{center}
  \medskip
} \fi

\bigskip
\begin{abstract}
Multivariate normal (MVN) probabilities arise in myriad applications, but they are analytically intractable and need to be evaluated via Monte-Carlo-based numerical integration. For the state-of-the-art minimax exponential tilting (MET) method, we show that the complexity of each of its components can be greatly reduced through an integrand parameterization that utilizes the sparse inverse Cholesky factor produced by the Vecchia approximation, whose approximation error is often negligible relative to the Monte-Carlo error. Based on this idea, we derive algorithms that can estimate MVN probabilities and sample from truncated MVN distributions in linear time (and that are easily parallelizable) at the same convergence or acceptance rate as MET, whose complexity is cubic in the dimension of the MVN probability. We showcase the advantages of our methods relative to existing approaches using several simulated examples. We also analyze a groundwater-contamination dataset with over twenty thousand censored measurements to demonstrate the scalability of our method for partially censored Gaussian-process models.
\end{abstract}

\noindent%
\vfill

\newpage
\spacingset{1.8} 

\section{Introduction \label{sec:intro}}

Multivariate normal (MVN) probabilities, which are integrals over MVN densities, must be evaluated in a variety of settings, most notably those involving Gaussian processes (GPs), including GP regression with inequality constraints \citep{da2012gaussian}, joint threshold-exceedance probabilities for GPs \citep{bolin2015excursion}, probit GP models for binary responses \citep{durante2019conjugate}, skew-normal models \citep{zhang2010spatial}, and censored scale-mixtures of normals for modeling tail dependence \citep{huser2017bridging}.
However, MVN probabilities are analytically intractable, and direct application of numerical integration rules often leads to highly inaccurate estimates \citep{genz1992numerical}. 

Current computational approaches for MVN probabilities are mostly based on the separation of variable (SOV) method \citep{genz1992numerical}. SOV applies a series of transformations such that the MVN probability can be written as an expectation of quantities with low variance, allowing an approximation of the expectation as the average of Monte Carlo (MC) samples. SOV consists of two stages: in the pre-processing stage, the Cholesky factor of the $n \times n$ MVN covariance matrix is computed in $\order(n^3)$ time; in the MC sampling stage, $N$ samples are drawn, each in $\order(n^2)$ time. The resulting SOV time complexity is $\order(n^3 + Nn^2)$. To improve upon the original SOV method,  \citet{gibson1994monte} introduced a univariate variable reordering technique used in the pre-processing stage that produces the Cholesky factor and further reduces the MC variance. Based on the series of integrand transformations used in the SOV method, \citet{botev2017normal} used importance sampling with minimax exponentially tilted (MET) normal proposal densities during the MC sampling stage to reduce variance. This results in substantially higher accuracy for tail MVN probabilities than SOV, but it requires iterative optimization of the proposal density in the pre-processing stage at a per-iteration complexity of $\order(n^3)$; {in practice, this optimization can be more expensive than the MC sampling for moderately large $n$ (e.g., $n > 1{,}000$)}. 
Sampling from truncated MVN (TMVN) distributions is closely connected to approximating MVN probabilities \citep{botev2017normal}. Specifically, the integrand in MET (or SOV) can be decomposed into two parts that can be viewed as a proposal density and a density ratio, which can be used for importance sampling from TMVN distributions at the same cost as estimating the corresponding MVN probability.
MET is currently the state-of-the-art in terms of achieving the highest accuracy for estimating MVN probabilities and the highest acceptance rate for sampling TMVN distributions.

\looseness=-1
For evaluating high-dimensional MVN probabilities with spatial covariance matrices, \citet{genton2018hierarchical} and \citet{cao2021exploiting} used hierarchical-matrix approximations to reduce the complexity of pre-processing and MC sampling for SOV by a factor of $n^{1/2}$. {\cite{bolin2015excursion} re-parameterized the SOV method with the inverse Cholesky factor to take advantage of the sparse precision matrix under Gaussian Markov random fields. However, they found this approach to often be computationally inefficient, because the sparsity was not well preserved during Cholesky factorization}.
Other recent approaches for spatial MVN probabilities have used versions of the Vecchia approximation \citep{Vecchia1988_abbrev}, which decomposes a multivariate density into a product of univariate conditional densities with small conditioning sets.
\citet{saha2022scalable} applied a Vecchia-type approximation to reduce the complexity of computing the Cholesky factor in the preprocessing stage, but its per-sample complexity remains as $\order(n^2)$. \citet{nascimento2022vecchia}  decomposed the $n$-dimensional MVN probability into a product of lower-dimensional MVN probability ratios by assuming conditional independence in the MVN cumulative distribution function (CDF); while this is based on a similar principle as the Vecchia approximation, it appears that the screening effect {(i.e., far-away information contributes little after conditioning on nearby points)} \citep{Stein2002} is weaker for MVN CDFs than for densities, resulting in substantially increased approximation error for CDFs {as will be shown empirically in this paper}.

Our proposed methods are also based on the Vecchia approximation. A Vecchia approximation of a MVN distribution results in a valid MVN distribution that has a covariance matrix with a sparse inverse Cholesky factor \citep[][]{Datta2016_abbrev,Katzfuss2017a}. Vecchia approximations can provide asymptotically exact inference in near-linear time for GPs with Mat\'ern kernels {subject} to boundary effects {and smoothness conditions} \citep{Schafer2020, kang2024asymptotic}{ --- such kernels and their relation to sparse precision matrices were also studied in \citet{Lindgren2011a}}.
Numerically, Vecchia approximations have been shown to be highly accurate even with small conditioning-set sizes in many settings, including spatial statistics and high-dimensional and multi-output GPs \citep[e.g.,][]{Datta2016_abbrev,Katzfuss2020,Schafer2020,Cao2023}. {Vecchia approximations can be applied to any covariance matrix, and we expect them to be most accurate if the covariance can be viewed as a stationary kernel with relatively small noise-to-signal ratio defined over some (potentially unknown) input space \citep{Kang2021}}.

We propose scalable evaluation of MVN probabilities and sampling from TMVN distributions in linear time by applying suitable Vecchia approximations to the MET approach. 
{Hence, we expect our methods to work well in the settings discussed above.}
We show that the dominant computations for MET and SOV are equivalent to calculating the conditional means and variances of the integration variables. 
We take full advantage of Vecchia's sparse inverse Cholesky factor when computing the transformed MET integrand as well as its gradient and Hessian, which allows us to optimize the MET proposal density and sample from TMVN distributions in linear time, an $\order(n^2)$ reduction compared to MET. We demonstrate theoretically that the Vecchia approximation error decreases exponentially with the number of nonzeros under certain regularity conditions. {Numerically, we show that under commonly used spatial kernels, the Vecchia error is often negligible relative to the Monte-Carlo error; for non-decaying kernels, such as constant correlation, the Vecchia approximation still provides a favorable trade-off between accuracy and computation cost}. 
We show that our proposed method facilitates model estimation and posterior inference for partially censored MVN models that offer substantial information gain over existing approaches, particularly in spatial applications involving censored responses.

The remainder of this document is organized as follows. In Section~\ref{sec:view}, we provide a review of the SOV and MET methods in a unified framework to set up the mathematical notation for our proposed method. In Section~\ref{sec:sampling}, we introduce our transformations of the SOV and the MET integrands to achieve linear complexity for MC sampling. In Section~\ref{sec:solve_gamma}, we derive the gradient and Hessian of our transformed MET integrand and show that solving for the proposal density used in MET can be achieved in $\order(n)$ time. In Section~\ref{sec:opt_tech}, {we introduce two optional techniques for improving accuracy, namely variable reordering and multi-level MC \citep{giles2008multilevel}}. Section~\ref{sec:numerical} provides numerical comparisons to existing methods for estimating MVN probabilities and sampling TMVN distributions. Section~\ref{sec:application} shows that our proposed method enables scalable estimation and inference for partially censored MVN models. Section~\ref{sec:conclusions} concludes. All proofs, a discussion on multivariate Student-$t$ (MVT) probabilities as well as variable reordering, and additional numerical results an can be found in the appendix. Our \texttt{R} package, 
\if1\blind{\texttt{VeccTMVN}}\fi \if0\blind {[blinded R package name]} \fi, 
which is available on CRAN, implements our methods, while code to reproduce our results can be found at\if1\blind
{
  \url{https://github.com/JCatwood/TMVN_Vecchia}.
} \fi 
\if0\blind
{
  [blinded Github repository].
} \fi

\section{Importance-Sampling Views of SOV and MET \label{sec:view}}

The MVN probability in an $n$-dimensional hyperrectangle has the general expression of:
\begin{align}
    \Phi_n(\ba, \bb; \bfmu, \bfSigma) = \int_{\ba}^{\bb} \phi_n(\bx; \bfmu, \bfSigma) \diff \bx = \int_{\ba}^{\bb} \frac{1}{(2\pi)^{n/2} |\bfSigma|^{1/2}} e^{-\frac{1}{2} (\bx - \bfmu)^\top \bfSigma^{-1} (\bx - \bfmu)} \diff \bx, \label{equ:MVN_prob_def} 
\end{align}
where $\bfmu$ and $\bfSigma$ are the mean and the covariance matrix of the MVN distribution, respectively; $\ba$ and $\bb$ are lower and upper integration limits defining the hyperrectangle; and $\phi_n(\cdot)$ is the MVN density function. 
Without loss of generality, $\bfmu$ is assumed zero and omitted throughout this paper, as $\Phi_n(\ba, \bb; \bfmu, \bfSigma) = \Phi_n(\ba - \bfmu, \bb - \bfmu; \bfzero, \bfSigma)$. 
In many applications, including spatial statistics and Gaussian processes, $\bfSigma$ is a kernel matrix whose $(i,j)$th entry $\bfSigma_{i, j} = \mathcal{K}(\bs_{i}, \bs_{j})$ is obtained by evaluating a covariance kernel $\mathcal{K}$ at the pair of locations or inputs $\bs_i$ and $\bs_j$ corresponding to the $i$th and $j$th entry of $\bx$.

Denoting the lower Cholesky factor of $\bfSigma$ by $\bL$, such that $\bfSigma = \bL \bL^{\top}$, a series of integrand transformations that start with the integration variable transformation $\bx = \bL \by$ may offer a clear overview of both SOV \citep{genz1992numerical} and MET \citep{botev2017normal}:
\begin{align}
    \Phi_n(\ba, \bb; \bfSigma) = {}
    &\int_{\tilde{a}_1}^{\tilde{b}_1} 
    \int_{\tilde{a}_2}^{\tilde{b}_2}
    \ldots
    \int_{\tilde{a}_n}^{\tilde{b}_n}
    \phi_n(\by; \bI_n) 
    \diff \by  \\
    ={} &\int_{\tilde{a}_1}^{\tilde{b}_1} \frac{\Phi(\tilde{b}_1) - \Phi(\tilde{a}_1)}{\Phi(\tilde{b}_1) - \Phi(\tilde{a}_1)} \phi(y_1)
    \cdots \int_{\tilde{a}_n}^{\tilde{b}_n} \frac{\Phi(\tilde{b}_n) - \Phi(\tilde{a}_n)}{\Phi(\tilde{b}_n) - \Phi(\tilde{a}_n)} \phi(y_n)
    \diff \by, \label{equ:SOV_transform} \\
    ={} &\int_{\tilde{a}_1}^{\tilde{b}_1} \frac{\Phi(\tilde{b}_1 - \gamma_1) - \Phi(\tilde{a}_1 - \gamma_1)}
    {\Phi(\tilde{b}_1 - \gamma_1) - \Phi(\tilde{a}_1 - \gamma_1)} 
    \frac{\phi(y_1)}{\phi(y_1 - \gamma_1)} \phi(y_1 - \gamma_1)
    \cdots \\
    &\int_{\tilde{a}_n}^{\tilde{b}_n} \frac{\Phi(\tilde{b}_n - \gamma_n) - \Phi(\tilde{a}_n - \gamma_n)}
    {\Phi(\tilde{b}_n - \gamma_n) - \Phi(\tilde{a}_n - \gamma_n)} 
    \frac{\phi(y_n)}{\phi(y_n - \gamma_n)} \phi(y_n - \gamma_n)
    \diff \by, \label{equ:botev_transform} \\
    ={} & E_{q^{B}(\by; \bfgamma)}[h^{B}(\by; \bfgamma)],\ \mbox{with}
\end{align}
\vspace{-3em}
\begin{align}
    \tilde{a}_i ={} &\frac{a_{i} - \sum_{j = 1}^{i - 1} L_{ij}y_j}{L_{ii}}, \quad 
    \tilde{b}_i = \frac{b_{i} - \sum_{j = 1}^{i - 1} L_{ij}y_j}{L_{ii}}, \\
    q^{B}(\by; \bfgamma) ={} &\prod_{i}^{n} \frac
    {\mathbbm{1}_{y_i \in (\tilde{a}_{i}, \tilde{b}_{i})}}
    {\Phi(\tilde{b}_i - \gamma_i) - \Phi(\tilde{a}_i - \gamma_i)} 
    \phi(y_i - \gamma_i),
    \label{equ:exp_tilt_proposal} \\
    h^{B}(\by; \bfgamma) = {}
    &\prod_{i}^{n}
    \left\{ \Phi(\tilde{b}_i - \gamma_i) - \Phi(\tilde{a}_i - \gamma_i) \right\}
    \frac{\phi(y_i)}{\phi(y_i - \gamma_i)}
    \mathbbm{1}_{y_i \in (\tilde{a}_{i}, \tilde{b}_{i})}. 
    \label{equ:exp_tilt_estimator}
\end{align}
where $\bfgamma$ are the parameters for the proposal density used in MET, and $\phi(\cdot)$ and $\Phi(\cdot)$ are the probability density function (PDF) and the cumulative probability function (CDF) of the univariate standard normal distribution, respectively. From the perspective of importance sampling, \eqref{equ:SOV_transform} and \eqref{equ:botev_transform} correspond to the SOV and the MET methods, respectively. Specifically, $q^{B}(\by; \bfgamma)$ is the proposal density and $h^{B}(\by; \bfgamma)$ is the density ratio between $\phi_n(\by; \bI_{n})$ truncated in $\ba \le \bL \by \le \bb$ and $q^{B}(\by; \bfgamma)$, up to a normalizing constant. MET samples $\by$ from $q^{B}(\by; \bfgamma)$ and then approximates $E_{q^{B}(\by; \bfgamma)}[h^{B}(\by; \bfgamma)]$ by the sample average.
Unlike TMVN distributions, $q^{B}(\by; \bfgamma)$ can be sampled as a sequence of $n$ univariate truncated normal distributions as implied by \eqref{equ:exp_tilt_proposal}. Hence, we can apply accept-reject sampling to TMVN distributions by using $q^{B}(\by; \bfgamma)$ as the proposal density to draw samples from $\phi_n(\bx; \bfSigma)$ truncated in $[\ba, \bb]$. SOV is a special case of MET when the exponential tilting parameters $\bfgamma = \bfzero$. Using
\begin{align}
    \hat{\bfgamma} = \argmin_{\bfgamma} \max_{\by: \, \ba \le \bL \by \le \bb} \log h^{B}(\by; \bfgamma), \label{equ:gamma_hat}
\end{align}
was proposed by \cite{botev2017normal} to minimize the worst likelihood ratio if we are to draw samples from the TMVN distribution. Solving for $\hat{\bfgamma}$ is discussed in Section~\ref{sec:solve_gamma}.

\section{Sampling at Linear Complexity \label{sec:sampling}}

\subsection{Conditional mean and variance under Vecchia}
\label{subsec:cond_mean_var_vecc}

After the pre-processing stage, the dominating $\order(n^2)$ complexity in the MC sampling of both SOV and MET comes from the computation of $\tilde{a}_i$ and $\tilde{b}_i$; specifically, for each $i=2,\ldots,n$, one must compute $\sum_{j = 1}^{i - 1} L_{ij}y_j$, which can be shown to be equivalent to $\mathrm{E}[x_i | \bx_{1 : i - 1}]$: 
\begin{proposition}
    \label{prp:cond_mean_var}
    Suppose $\bx \sim \phi_{n}(\bx; \bfSigma)$, $\bL$ is the lower Cholesky factor of $\bfSigma$, and $\bx = \bL \by$. Then
    \begin{align}
        E[x_i | \bx_{1 : i - 1}] = \sum_{j = 1}^{i - 1} L_{ij}y_j, \quad \mathrm{Var}[x_i | \bx_{1 : i - 1}] = L_{i, i}^{2},\ \mbox{where }\bx_{i:j} = [x_i,x_{i+1},\ldots,x_j]^\top.
    \end{align}
\end{proposition}
{Proposition~\ref{prp:cond_mean_var} is a well-known result, whose proof is omitted in this paper}. Each conditional expectation and variance can be computed at sublinear complexity via the Vecchia approximation \citep{Vecchia1988_abbrev}, which truncates the conditioning sets in the conditional densities to subsets of sizes no bigger than $m$:
\begin{align}
    f(\bx) = f(x_1) \cdot f(x_2 | x_1) \cdots f(x_n | \bx_{1 : n - 1}) 
    \approx f(x_1) \cdot f(x_2 | x_{c(1)}) \cdots f(x_n | \bx_{c(n)}), 
\end{align}
where $f(\cdot)$ is a generic density and the conditioning set $c(i)$ is a subset of $\{1, \ldots, i - 1\}$ of size $\min(m, i - 1)$. When $f$ is the normal density, the Vecchia approximation amounts to
\begin{align}
    \phi_{n}(\bx; \bfSigma) \approx \phi(x_1; \eta_{1}, l_{1}^{2}) \cdot \phi(x_2 ; \eta_{2}, l_{2}^{2}) \cdots \phi(x_n ; \eta_{n}, l_{n}^{2}), 
\end{align}
where $\eta_{i} = E[x_i | \bx_{c(i)}]$ and $l_{i}^{2} = \mbox{Var}[x_i | \bx_{c(i)}]$ can be computed straightforwardly based on an $(m+1) \times (m+1)$ submatrix of $\bfSigma$. Each conditioning set is typically chosen as the indices of the $m$ nearest previously ordered neighbors. For covariances $\bfSigma_{i, j} = \mathcal{K}(\bs_{i}, \bs_{j})$ based on isotropic kernels, Euclidean distance between inputs $\bs_i$ and $\bs_j$ can be used to define nearest neighbors; for other covariance matrices, the correlation distance $(1-|\rho_{ij}|)^{1/2}$ with $\rho_{ij} = \bfSigma_{ij}/(\bfSigma_{ii}\bfSigma_{jj})^{1/2}$ implicitly selects neighbors in a suitable transformed input space \citep{Kang2021}.
The Vecchia approximation is based on the screening effect \citep{Stein2002}, which describes the phenomenon that the marginal information gain from far-away responses is often minimal after conditioning on nearby responses. 
\citet{Schafer2020} showed that, for each $i=1,\ldots,n$, the nonzero entries in the $i$th column of the inverse Cholesky factor $\bV = \bL^{-\top}$ can be obtained as:
\begin{equation}
    \bV_{\bar{c}(i), i} = \bu_i \bu_{i,1}^{-1/2}, \qquad \text{with } \bu_i = (\bfSigma_{\bar{c}(i), \bar{c}(i)})^{-1} \be_1, \label{equ:inv_chol_formula}
\end{equation}
where $\bar{c}(i) = [i, c(i)^{\top}]^{\top}$ and $\be_1$ is a vector whose first entry is 1 and all other entries are zero. Hence, $\bV$ can be computed using $\order(nm^3)$ operations and in parallel. 
Numerical evidence that a small $m \leq 50$ is typically sufficient for many spatial and GP settings with very large $n$ is also abundant in the literature (see Section \ref{sec:intro} for references). Therefore, in such settings, we can approximately view $m$ as fixed as $n$ grows.
The conditional expectation and variance in Proposition \ref{prp:cond_mean_var} can be computed in $\order(m)$ time for each $i$ when applying the Vecchia approximation to the covariance, resulting in a per-sample complexity of $\order(nm)$.
\begin{proposition}
    \label{prp:cond_mean_Vecc}
    Suppose $\bx \sim \phi_{n}(\bx; \bfSigma)$, where $\bfSigma$ is the covariance matrix under the Vecchia approximation defined by the conditioning sets $\{c(i)\}_{i = 1}^{n}$. Then,
    \begin{align}
        \eta_{i} &\defeq E[x_i | \bx_{1 : i - 1}] = \bA_{i, :} \bx ,  \label{equ:cond_mean_vecc} \\
        l_{i}^{2} &\defeq \mbox{Var}[x_i | \bx_{1 : i - 1}] = \bV_{i, i}^{-2},
    \end{align}
    where $\bA$ is a sparse matrix $\bA^{\top} = (\diag(\bl)^{-1} - \bV) \cdot \diag(\bl)$, $\bV$ is the (sparse) inverse Cholesky factor of $\bfSigma$, $\bfSigma = (\bV \bV^{\top})^{-1}$, and $\bl$ is the inverse of the diagonal entries of $\bV$.
\end{proposition}
\noindent
{Proposition~\ref{prp:cond_mean_Vecc} is similar to existing results \citep[e.g.,][Thm.~2.3]{rue2005gaussian}}. Its proof is included in the appendix for completeness. The definition of $\bl$ in  Proposition~\ref{prp:cond_mean_Vecc} coincides with that above, namely the univariate conditional standard deviations. {Notice that our definitions of $\bA$ and $\bl$ share similarities with \cite{bolin2015excursion}, which re-parameterized the SOV method with the inverse Cholesky factor, but \cite{bolin2015excursion} computed $\bV$ from the sparse precision matrix, hence leading to higher complexity and less sparsity in $\bV$. The conditional expectations and variances computed as proposed in Proposition~\ref{prp:cond_mean_Vecc} result in biased MVN probabilities. However, the bias is shown in Proposition~\ref{prp:vecchia_rel_err} to decrease exponentially with the number of non-zero entries in $\bV$ under certain regularity conditions, {covering Mat\'ern kernels with smoothness $\nu$ such that $\nu + d/2$ is an integer, where $d$ is the dimension of the domain over which Mat\'ern kernels are defined \citep{Schafer2020}.}
\begin{proposition}[informal version]
    \label{prp:vecchia_rel_err}
    Let $\bfSigma$ and $\bfSigma_{V}$ denote the original covariance matrix and the covariance matrix under the Vecchia approximation, respectively. Under conditions specified in the appendix, the relative error
    \begin{align}
        \frac{|\Phi(\ba, \bb; \bfSigma) - \Phi(\ba, \bb; \bfSigma_{V})|}{\Phi(\ba, \bb; \bfSigma)}
    \end{align}
    decreases exponentially with the number of nonzero entries in $\bV$.
\end{proposition}
The full version of Proposition~\ref{prp:vecchia_rel_err} and its proof can be found in the appendix.}

\subsection{Transforming SOV and MET}

With Proposition~\ref{prp:cond_mean_Vecc}, we transform the SOV and MET algorithms such that MC sampling can be performed at linear complexity as described in Algorithm \ref{alg:exp_tilt_vecchia}. The new method is coined Vecchia MET (VMET). The exponential tilting parameter $\bfgamma$ can be obtained as discussed in Section~\ref{sec:solve_gamma} during the pre-processing stage, or it can be set as $\bfgamma = \bfzero$ to obtain Vecchia SOV (VSOV) as a special case. {VMET uses the original integration variable $\bx$, and hence we define $q(\bx) = q^{B}(\bL^{-1} \bx)$ and $h(\bx) = h^{B}(\bL^{-1} \bx)$ as the proposal density and the density ratio, up to a normalizing constant, used in VMET.}
\begin{algorithm}[h!]
\spacingset{1}
\caption{Vecchia MET integrand}
\label{alg:exp_tilt_vecchia}
\KwInput{$\ba, \bb, \bA, \bl, \bfgamma, \bw$}
\KwResult{One sample of the MET integrand and (optionally) the proposal distribution}
\begin{algorithmic}[1]
\STATE Initialize $\bx \gets \bfzero$, $h \gets 1$
\FOR{$i = 1, 2, \ldots, n$}
  \STATE if $i > 1$, $\mu_{i} \gets \bA_{i, :} \bx$\ \ else $\mu_{i} \gets 0$ \label{stp:cond_mean_vecc}
 \STATE {\color{black}$\tilde{a}_{i} \gets \frac{a_{i} - \mu_{i}}{l_{i}} - \gamma_i$, $\tilde{b}_{i} \gets \frac{b_{i} - \mu_{i}}{l_{i}} - \gamma_i$,} $p_{i} \gets \Phi(\tilde{b}_{i}) - \Phi(\tilde{a}_{i})$
 \STATE {\color{black}$y_{i} \gets \Phi^{-1}(w_{i} \cdot p_{i} + \Phi(\tilde{a}_{i})) + 
 \gamma_{i}$, $x_i \gets \mu_{i} + y_{i} \cdot l_{i}$}
 \STATE {\color{black}$h \gets h \cdot p_{i} \cdot \phi(y_{i}) / \phi(y_{i} - \gamma_i)$}
\ENDFOR
\IF{Sampling from TMVN}
  \RETURN $h$ and $\bx$
\ELSE
  \RETURN $h$
\ENDIF
\end{algorithmic}
\end{algorithm}
We use $\bw$ to denote a sample from the unit hypercube in $\mathbb{R}^n$, offering the flexibility of choosing different MC rules. Compared to MET, the complexity of each iteration (i.e., for each $i$) is reduced from $\order(n)$ to $\order(m)$ employing the sparsity of $\bA$ in Step~\ref{stp:cond_mean_vecc}. Therefore, the per-sample complexity is reduced from $\order(n^2)$ to $\order(n)$ during the MC sampling stage. We refer to Algorithm~\ref{alg:exp_tilt_vecchia} as the `integrand' to distinguish from the `method', which includes both pre-processing and MC sampling. 
Algorithm~\ref{alg:exp_tilt_vecchia} can be also used for accept-reject sampling of TMVN distributions, where $\bx$ is a sample generated from the proposal density $q(\bx)$ and $h$ is the PDF ratio. VMET allows us to fully exploit the sparseness of $\bV$ by using the original integration variable $\bx$. Since the Vecchia approximation amounts to a valid MVN density with (slightly) different covariance, Algorithm~\ref{alg:exp_tilt_vecchia} enjoys the same analytical properties and efficacy in the variance reduction of the integrand as the original MET method \citep{botev2017normal}. This is supported by our theoretical and numerical results in Sections~\ref{sec:solve_gamma} and \ref{sec:numerical}, respectively. { MVT probabilities can be defined as scale mixtures of MVN probabilities. Based on this, we provide modifications of the proposed VMET integrand for MVT probabilities, along with associated numerical results, in Appendix~\ref{sec:appendix_MVT}.}


\section{Solving For Exponential Tilting Parameters}
\label{sec:solve_gamma}

\citet{botev2017normal} proposed to find the optimal $\bfgamma$, denoted by $\hat{\bfgamma}$, by minimizing the worst likelihood ratio as shown in \eqref{equ:gamma_hat}. Denoting $\log h$ by $\psi$, \citet{botev2017normal} showed that $\psi$ is convex with respect to (w.r.t.) $\bfgamma$ and concave w.r.t.\ $\by$. Hence, finding $\hat{\bfgamma}$ amounts to finding the solution $(\hat{\by}, \hat{\bfgamma})$ of the nonlinear system $\nabla \psi = \bfzero$ if the solution satisfies $\ba \le \bL \hat{\by} \le \bb$, which was empirically shown to be true by \cite{botev2017normal}. Under our parameterization of $\psi$ that uses $\bx$ instead of $\by$, it can be shown that the convex-concave property still holds.
\begin{proposition}
    \label{prp:convex_concave}
    For fixed $\bx$, $\psi$ is a convex function of $\bfgamma$; for fixed $\bfgamma$, $\psi$ is a concave function of $\bx$. Therefore, 
    \[
        \hat{\bfgamma} = \argmin \max_{\ba \le \bx \le \bb} \psi(\bx; \bfgamma)
    \]
    is a saddle point problem with a unique solution given by $\nabla \psi(\bx; \bfgamma) = \bfzero$.
\end{proposition}
\noindent

However, solving $\nabla \psi = \bfzero$ amounts to finding the solution of a challenging non-linear system of $2n$ variables. \cite{botev2017normal} used the trust-region method described in \citet{powell1970hybrid}, which minimizes $g \defeq \|\nabla \psi\|^2$ through building a quadratic approximation of $g$ at the current values of $(\by, \bfgamma)$. Notice that \cite{botev2017normal} viewed $\psi$, hence $g$, as functions of $(\by, \bfgamma)$ where as we view them as functions of $(\bx, \bfgamma)$. The quadratic approximation of $g$ requires the computation of $\bH \cdot \nabla \psi$ and $\bH^{-1} \cdot \nabla \psi$, where $\bH$ denotes the Hessian matrix of $\psi$. {This involves expensive $\order(n^3)$ matrix operations and renders MET less scalable than other methods}.

In Proposition~\ref{prp:psi_grad_hession}, we derive $\nabla \psi$ and $\bH$ under the parameterization of $\psi$ that uses $\bx$, $\bA$ and $\bl$. Utilizing the sparse-matrix representation of $\bH$, we propose an $\order(n)$ gradient-based method for solving $\hat{\bfgamma}$ that proves to be significantly more efficient than and as effective as using the trust-region method in \citet{botev2017normal}.
\begin{proposition}
    \label{prp:psi_grad_hession}
    Denote the logarithm of the density ratio, $\log h(\bx; \bfgamma)$, from \eqref{equ:exp_tilt_estimator} by $\psi$. Under the Vecchia approximation, the gradient and Hessian of $\psi$ can be derived as:
\begin{align}
    &\frac{\partial \psi}{\partial \bx} 
    ={} 
    - (\bI - \bA)^{\top} \bD_{\bl}^{-1} \bfgamma + \bA^{\top} \bD_{\bl}^{-1} \bfPsi, \quad
    \frac{\partial \psi}{\partial \bfgamma} = {} 
    \bfgamma - \bD_{\bl}^{-1} (\bx - \bfmu_{c}) + \bfPsi,
    \label{equ:grad} \\
    &\bH =
    \begin{bmatrix}
        \bA^{\top} \bD_{\bl}^{-1} \bfPsi^{'} \bD_{\bl}^{-1} \bA & -(\bI - \bA)^{\top} \bD_{\bl}^{-1} + \bA^{\top} \bD_{\bl}^{-1} \bfPsi^{'} \\
        -\bD_{\bl}^{-1} (\bI - \bA) + \bD_{\bl}^{-1} \bfPsi^{'} \bA & \bI + \bfPsi^{'}
    \end{bmatrix},
    \label{equ:hessian}
\end{align}
where $\bD_{\bl}$ is the $n \times n$ diagonal matrix with diagonal entries equal to $\bl$, $\bfPsi$ is a vector of length $n$ with $i$th entry
\[
    \Psi_{i} = 
    \frac
    {\phi(\tilde{a}_{i}; \gamma_i, 1) - \phi(\tilde{b}_{i}; \gamma_i, 1)}
    {\Phi(\tilde{b}_i - \gamma_{i}) - \Phi(\tilde{a}_i - \gamma_{i})}, 
\]
and $\bfPsi'$ is a $n \times n$ diagonal matrix with $i$th diagonal entry
\[
    \Psi_{i, i}^{'} = 
    \frac
    {(\tilde{a}_{i} - \gamma_i)\phi(\tilde{a}_{i}; \gamma_i, 1) - (\tilde{b}_{i} - \gamma_i)\phi(\tilde{b}_{i}; \gamma_i, 1)}
    {\Phi(\tilde{b}_i - \gamma_{i}) - \Phi(\tilde{a}_i - \gamma_{i})} -
    \Psi_{i}^{2},
\]
with $\tilde{\ba} = \bD_{\bl}^{-1} (\ba - \bA\bx)$ and $\tilde{\bb} = \bD_{\bl}^{-1} (\bb - \bA\bx)$.
\end{proposition}

Based on these expressions, we propose to minimize $g$ for VMET using a gradient-based optimizer, for which $\nabla g = \bH \cdot \nabla \psi$ can be computed efficiently in $\order(n)$ time, and which does not require the more expensive $\bH^{-1} \cdot \nabla \psi$.
Specifically, due to the sparsity of $\bA$ with fewer than $nm$ non-zero coefficients, $\nabla \psi$ in \eqref{equ:grad} and matrix-vector products involving $\bH$ can be computed in $\order(nm)$ time. 
In our implementation, we use the limited-memory Broyden-Fletcher-Goldfarb-Shanno (L-BFGS) algorithm \citep{liu1989limited} as our optimizer, which creates a low-rank approximation of the Hessian of $g$, maintaining an overall $\order(n)$ complexity. Another advantage of VMET over MET is that the Hessian of $\psi$ with respect to $\bx$, $\frac{\partial^{2} \psi}{\partial \bx^{2}}$, can be represented using the sparse matrix $\bA$, and so  $\argmax_{\bx} \psi(\bx; \bfgamma)$ can be solved efficiently with second-order optimization algorithms. Finding $\max_{\bx} \psi(\bx; \bfgamma)$, which is the logarithm of the maximum PDF ratio (i.e., $\max \log h$ in \eqref{equ:exp_tilt_estimator}), accurately for a given $\bfgamma$ is crucial to the accept-reject sampling of TMVN distributions introduced in Section~\ref{sec:view}. 

An important property of the MET integrand is the vanishing relative error property (VRE) that was proved in \cite{botev2017normal} based on the minimax definition of $\hat{\bfgamma}$. The VRE property can also be proved for our proposed VMET integrand.
\begin{proposition}
    \label{prp:VRE}
    The integrand $h(\bx; \hat{\bfgamma}) = \exp \psi(\bx; \hat{\bfgamma})$ has vanishing relative error:
    \[
        \limsup_{\alpha \rightarrow + \infty} \frac{\mbox{var}(h(\bx; \hat{\bfgamma}))}{h^{2}(\bx; \hat{\bfgamma})} = 0,
    \]
    when the integration limits are $(\alpha \ba, +\bm{\infty})$ and $a_{i} > 0$ for $i = 1, \ldots, n$. Here, the variance is taken with respect to $\bx$ sampled from $q(\bx; \hat{\bfgamma})$.
\end{proposition}

\noindent
The VRE property suggests that the proposal density $q(\bx; \hat{\bfgamma})$ can become indistinguishable from the MVN density $\phi_n(\bx; \bfSigma)$ truncated in the region of $(\alpha \ba, +\bm{\infty})$.


\section{{Optional Techniques for Higher Accuracy}}
\label{sec:opt_tech}

\subsection{Variable Reordering Based on Vecchia}
\label{sec:var_reorder}

An integration-variable reordering method, referred to as univariate variable reordering was described in \citet{gibson1994monte} to empirically improve the MC sampling variance. Univariate reordering, provided in Algorithm \ref{alg:univar_reorder} in the appendix for completeness, iteratively computes conditional univariate normal probabilities and prioritizes the one with the smallest probability, resulting in $\order(n^3)$ complexity.
We propose a fast Vecchia-based version of the univariate variable reordering in Algorithm~\ref{alg:vecc_reorder}. 
Under the Vecchia approximation, the conditional expectation and variance in Line~\ref{stp:cond_mean_var} of Algorithm~\ref{alg:univar_reorder} can be computed in $\order(m^3)$ time for each $j$ as indicated by Line~\ref{stp:cond_mean_var_vecc} of Algorithm~\ref{alg:vecc_reorder}, amounting to a total algorithm complexity of $\order(n^2)$. Because the conditioning sets in the Vecchia approximation are subsets of previous indices (i.e., $c(i) \subset \{1, \ldots, i - 1\}$), we need to construct and update the conditioning sets during univariate reordering as described in Lines~\ref{stp:cond_set_update_1} and \ref{stp:cond_set_update_2} of Algorithm~\ref{alg:vecc_reorder}, where the conditioning sets are chosen as the $m$ previously indexed integration variables that have the strongest correlation with the $j$-th integration variable.

Variable reordering can be reduced to linear complexity under a special case of the Vecchia approximation with $c(i) = \{1,\ldots,m\}$ for $i>m$, which results in the fully independent conditional (FIC) approach \citep{Snelson2007,Katzfuss2017a}, a popular GP approximation in machine learning. 
Under the FIC assumption, the variable ordering (after selecting the first $m$ variables) can be obtained in a single iteration, hence achieving an overall $\order(n)$ complexity. For the first $m$ variables, we can simply run the loop in Algorithm~\ref{alg:vecc_reorder} for $m$ iterations. We refer to this $\order(n)$ ordering as the FIC-based variable reordering to distinguish it from the $\order(n^2)$ Vecchia-based variable reordering in Algorithm~\ref{alg:vecc_reorder}. We provide a numerical comparison of the univariate reordering, the Vecchia-based variable reordering, and the FIC-based variable reordering in Section~\ref{subsec:var_reorder}.

Algorithm~\ref{alg:VMET} summarizes our proposed method for estimating MVN probabilities and sampling TMVN distributions by combining the results from Sections~\ref{sec:sampling} and \ref{sec:solve_gamma} with the variable reordering introduced in this section.
\begin{algorithm}[h!]
\spacingset{1}
\caption{The VMET method}
\label{alg:VMET}
\KwInput{$n, \ba, \bb, \bfSigma, N$}
\KwResult{MVN probability estimate \textbf{OR} samples from a TMVN distribution}
\begin{algorithmic}[1]
\IF{Reordering needed}
\STATE $\br \gets \mbox{Algorithm~\ref{alg:vecc_reorder}}(n, \ba, \bb, \bfSigma)$, $\ba \gets \ba[\br]$, $\bb \gets \ba[\br]$, $\bfSigma \gets \bfSigma[\br, \br]$
\ENDIF
\STATE Compute $\bA$ and $\bl$ defined in Section~\ref{subsec:cond_mean_var_vecc} with $\bfSigma$
\STATE Minimize (e.g., using L-BFGS) $g = \|\nabla \psi\|^{2}$ described in \eqref{equ:grad} with $\nabla g = \bH \cdot \nabla \psi$ described in \eqref{equ:grad} and \eqref{equ:hessian}. Denote the solution by $(\hat{\bx}, \hat{\bfgamma})^{\top}$
\FOR{$i = 1, \ldots, N$}
\STATE Draw $\bw$ from the unit hypercube. Run Algorithm~\ref{alg:exp_tilt_vecchia}($\ba, \bb, \bA, \bl, \bfgamma, \bw$) 
\IF{Samples from a TMVN distribution is needed}
\STATE Generate $U \sim \mbox{uniform}(0, 1)$, if $U \exp \psi(\hat{\bx}, \hat{\bfgamma}) < p$, accept $\bx$, reject otherwise
\ELSE
\STATE Record $p_i = p$. The average of $\{p_{i}\}$ is the estimate of $\Phi_{n}(\ba, \bb, \bfSigma)$
\ENDIF 
\ENDFOR
\end{algorithmic}
\end{algorithm}
In most settings, $\bfSigma$ is a kernel matrix that does not need to be stored explicitly and whose entries can be computed as needed based on the kernel and the relevant inputs or locations, resulting in a $\order(n)$ memory footprint. 
After variable reordering, VMET has a time complexity of $\order(Tnm + Nnm)$, where the summands correspond to optimizing the exponential tilting parameter $\bfgamma$ and drawing MC samples of the VMET integrand, respectively. The number of iterations $T$ for optimizing $\bfgamma$ typically ranges between $500$ and $1{,}000$, while the MC sample size $N$ is typically between $10^4$ and $10^5$. Although Vecchia-based variable reordering has quadratic complexity in $n$, it is usually not needed repeatedly for estimating the unknown parameters in $\bfSigma$ to achieve a smooth likelihood surface that is beneficial to optimization, as will be discussed in Section~\ref{subsec:parm_est}. 

\subsection{{Multi-level Monte Carlo sampling}}

{ Multi-level MC \citep{giles2008multilevel} is a generic sampling technique that can be applied in our setting to mitigate the bias of the proposed VMET integrand. Specifically, multi-level MC draws $N_{1}$ low-cost-low-accuracy MC samples and $N_{2}$ high-cost-high-accuracy MC samples, with $N_{2} \ll N_{1}$, where the high-cost samples are correlated with (a subset of) the low-cost ones. The averaged difference between the two groups provides a bias estimate. Let $\{h_{l}^{m_{1}}\}_{l = 1}^{N_{1}}$ and $\{h_{l}^{m_{2}}\}_{l = 1}^{N_{2}}$ denote the samples of $h$ from \eqref{equ:exp_tilt_estimator} drawn at $m = m_{1}$ and $m = m_{2}$, respectively, with $m_{1} < m_{2}$. An estimator for the bias is
\begin{align}
    \hat{\epsilon} = \frac{1}{N_{2}} \left( 
    \sum_{l = 1}^{N_{2}} h_{l}^{m_{1}} - \sum_{l = 1}^{N_{2}} h_{l}^{m_{2}}
    \right),
\end{align}
where $h_{l}^{m_{1}}$ and $h_{l}^{m_{2}}$ are correlated for each $l=1,\ldots,N_2$ by using the same randomness profile to reduce the impact from the variance of $h$. The resulting multi-level-MC probability estimate is $\hat{h} = \frac{1}{N_{1}} \sum_{l = 1}^{N_{1}} h_{l}^{m_{1}} - \hat{\epsilon}$. The dependence between $\frac{1}{N_{1}} \sum_{l = 1}^{N_{1}} h_{l}^{m_{1}}$ and $\hat{\epsilon}$ is minimal, because $N_{2} \ll N_{1}$. Multi-level MC heuristically improves the accuracy of the VMET method, but its practical performance may depend on the distribution of $h_{l}^{m_{1}} - h_{l}^{m_{2}}$. In certain cases, the bias of the VMET integrand may not monotonically decrease with $m$, which was observed in our numerical simulations (e.g., see Figures~\ref{fig:low_dim_exp} and \ref{fig:high_dim_exp}). This may render multi-level MC less effective in the bias reduction, because high-cost samples may not result in smaller bias and hence, we categorize multi-level MC as an optional technique and demonstrate its impact under a constant correlation structure in Section~\ref{subsec:const_corr_sim}.}

\section{Numerical comparison \label{sec:numerical}}

In this section, we first show that integration variable reordering has a significant impact on the variance of the MET and VMET integrand samples when evaluating MVN probabilities with spatial covariance matrices. Next, we compare the following methods:
\begin{description}
    \spacingset{1}
    \item[VMET:] our proposed Vecchia-based method, summarized in Algorithm \ref{alg:VMET}
    \item[SOV:] the SOV method from \citet{genz1992numerical, genz2009computation}
    \item[MET:] the MET method from \citet{botev2017normal}
    \item[TLR:] the tile-low-rank (TLR) SOV method from \citet{cao2021exploiting}
    \item[VCDF:] the Vecchia approximation of MVN CDFs from \citet{nascimento2022vecchia}
\end{description}
\vspace{-1em}
The original implementation of SOV in the R package \texttt{mvtnorm} \citep{genz2009computation} only allows input MVN dimensions $n$ up to $1{,}000$; for $n>1{,}000$ we used the SOV implementation in \texttt{tlrmvnmvt} \citep{cao2022tlrmvnmvt}. The implementations used for MET, TLR, and VCDF were the \texttt{TruncatedNormal} \citep{botev2017normal} R package, the \texttt{tlrmvnmvt} \citep{cao2022tlrmvnmvt} R package, and the R package from \url{https://github.com/Recca2012/CDFApprox}, respectively. {\texttt{tlrmvnmvt}, \texttt{TruncatedNormal}, and our implemented VMET all utilized \texttt{C/C++} source code through \texttt{Rcpp} \citep{eddelbuettel2011rcpp}, while VCDF is based on a pre-compiled binary package}. Computation times were measured on a standard scientific workstation.

\subsection{Variable reordering comparison}
\label{subsec:var_reorder}

\looseness=-1
In this section, we show that the variable reordering heuristic proposed for SOV \citep{gibson1994monte} can also improve the results for MET and VMET. We demonstrate that our proposed Vecchia-based variable reordering (Algorithm~\ref{alg:vecc_reorder}) can achieve the same level of improvement in reducing the MC variance as the univariate reordering (Algorithm~\ref{alg:univar_reorder}). SOV and TLR are not considered in this section, because their benefits from variable reordering have been documented in the literature \citep[e.g.,][]{genz2009computation, cao2022tlrmvnmvt}. 

We considered three low-dimensional ($n = 900$) MVN probabilities to allow comparisons to the non-scalable MET method as a gold standard. All three examples used a Mat\'ern-$1.5$ kernel in $\mathbb{R}^2$ with variance $1$, range $0.1$, and nugget variance $0.01$. The three scenarios, which are summarized in Table~\ref{tbl:low_dim_exp}, cover a variety of settings, including regular versus irregular locations, same versus different integration intervals, and tail versus centered probabilities.
\vspace{-1.7em}
\begin{table*}[h!]
    \centering
    \spacingset{1}
    \begin{tabular}{c|c|c|c}
         & spatial locations $\{\bs_{i}\}_{i = 1}^{n}$ &lower limits $\ba$ & upper limits $\bb$ \\
         \hline
         Scenario 1 & grid in $[0, 1]^2$ &  $- \bm{\infty}$ & $\bfzero$ \\
         Scenario 2 & Latin hypercube in $[0, 1]^2$ & $- \bm{\infty}$ & $\mbox{Uniform}(-2, 0)$ \\
         Scenario 3 & grid in $[0, 1]^2$ &  $- \mathbf{1}$ & $\mathbf{1}$ 
    \end{tabular}
    \caption{Spatial locations and integration limits for three simulation scenarios. $\mbox{Uniform}(-2, 0)$ refers to generating random numbers from the uniform distribution in $(-2, 0)$ independently for each coefficient.}
    \label{tbl:low_dim_exp}
\end{table*}
\vspace{-1em}

Figure~\ref{fig:order_cmp_MET_VMET} in the appendix compares the MET and VMET (with $m = 30$) integrands with and without univariate variable reordering. 
Variable reordering reduced the variability of both integrands, hence improving the accuracy for MVN probability estimation. The effectiveness of variable reordering varied across different MVN probabilities. In Scenario 3, where integration limits were centered and the same, variable reordering had a smaller impact than in the other two cases. Furthermore, our proposed VMET integrand had overall negligible bias and the same level of variability compared with the MET integrand.

\begin{figure*}[h!]
\centering
	\begin{subfigure}{.33\textwidth}
	\centering
 	\includegraphics[width =.99\linewidth]{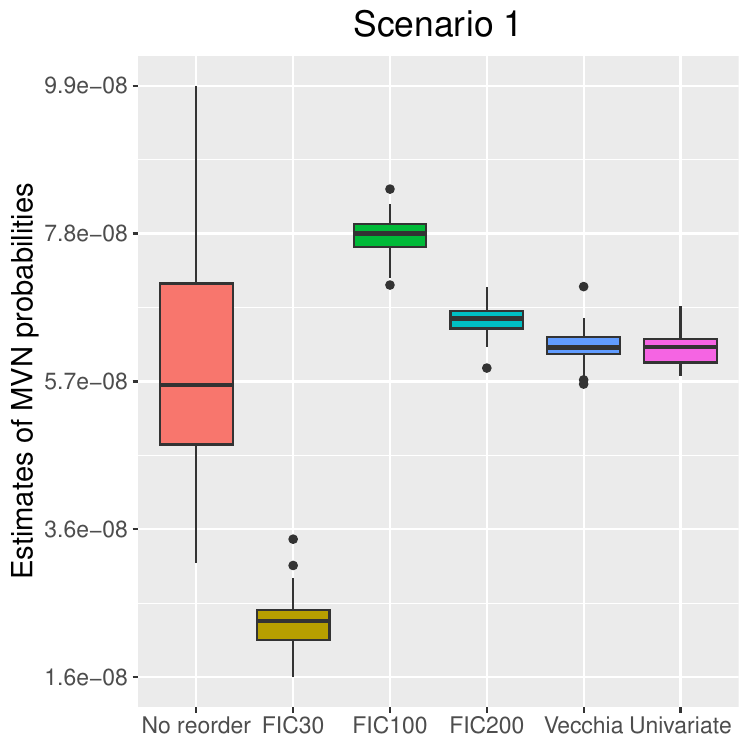}
	\end{subfigure}%
\hfill
	\begin{subfigure}{.33\textwidth}
	\centering
 	\includegraphics[width =.99\linewidth]{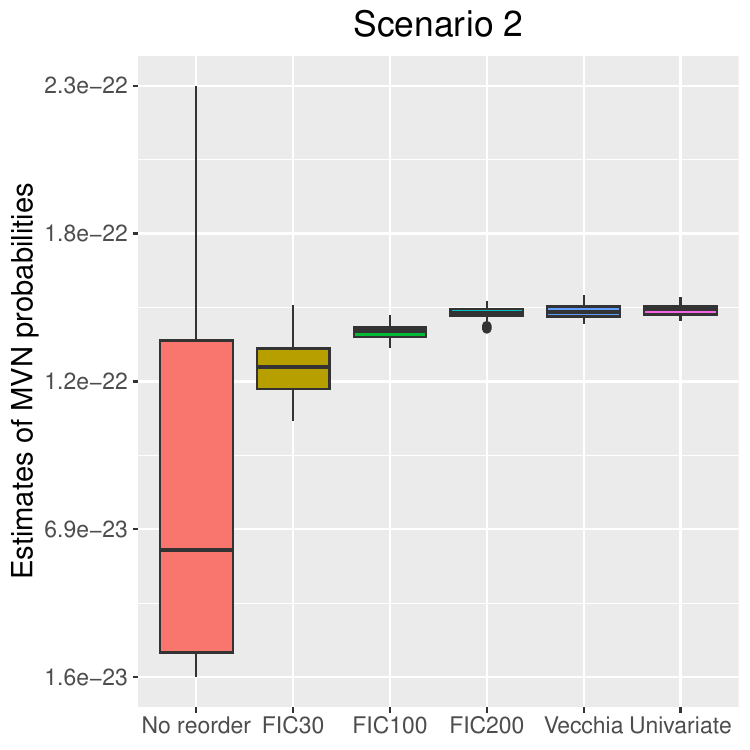}
	\end{subfigure}%
\hfill
	\begin{subfigure}{.33\textwidth}
	\centering
 	\includegraphics[width =.99\linewidth]{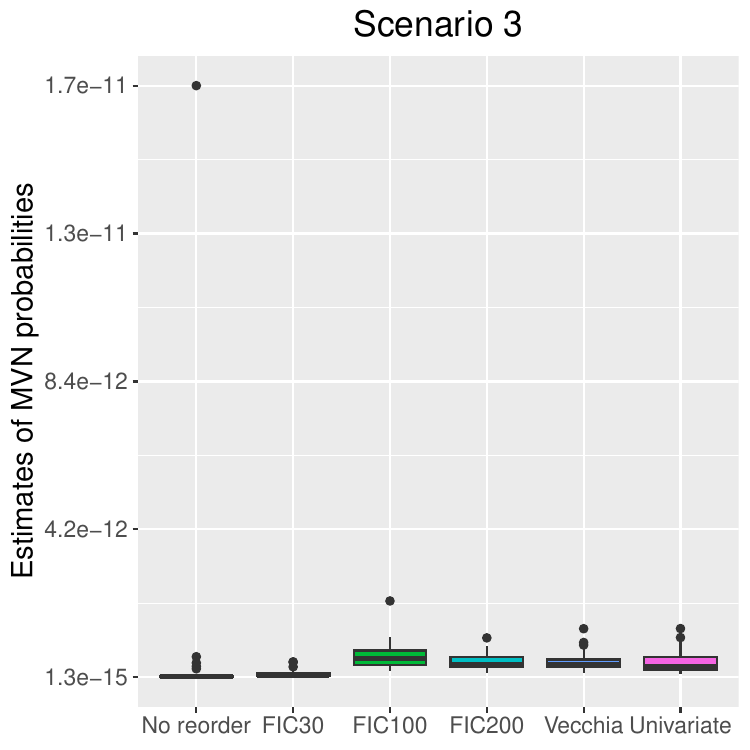}
	\end{subfigure}%
  \caption{Comparison of FIC-based variable reordering ($m = 30$, $100$, $200$), Vecchia-based variable reordering ($m = 30$), and univariate reordering for the three scenarios described in Table \ref{tbl:low_dim_exp}. Each boxplot consists of $30$ estimates using the VMET integrand.}
\label{fig:order_cmp_FIC_Vecc}
\end{figure*}

Figure~\ref{fig:order_cmp_FIC_Vecc} compares the VMET integrand based on four different reordering approaches: no reordering, FIC-based reordering (last paragraph of Section~\ref{sec:var_reorder}), Vecchia-based reordering, and classic univariate reordering. 
We used conditioning-set size $m = 30$ for Vecchia and $m = 30, 100, 200$ for FIC to showcase the convergence of the FIC-based variable reordering to the Vecchia based variable reordering. The improvement from the FIC-based reordering depended on the experiment scenario and the choice of $m$, but FIC-based reordering already achieved significant variance reduction with $m = 100$ and similar level of variance reduction as Vecchia-based and univariate reorderings with $m = 200$. Furthermore, when the FIC approximation has been applied to the underlying GP, the FIC-based reordering will produce the same ordering as the univariate reordering. Vecchia-based reordering and univariate reordering were almost identical in MC variance reduction, and both were more effective than the FIC-based reordering with $m \le 100$. Our proposed Vecchia-based reordering reduces the computation cost by a factor of $n$ compared to univariate reordering.

\subsection{Comparison for low-dimensional MVN probabilities}
\label{subsec:low_dim_numerical}

\begin{figure*}[h!]
    \centering
    \begin{subfigure}{.3\textwidth}
    \centering
    \includegraphics[width =.99\linewidth, trim=0 0 1in 0, clip]{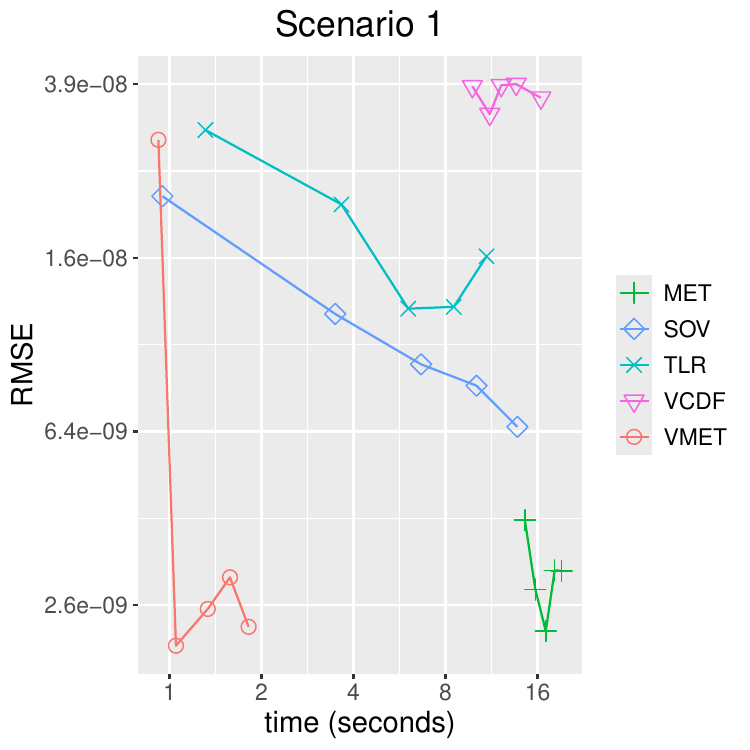}
    \end{subfigure}%
    \begin{subfigure}{.3\textwidth}
    \centering
    \includegraphics[width =.99\linewidth, trim=0 0 1in 0, clip]{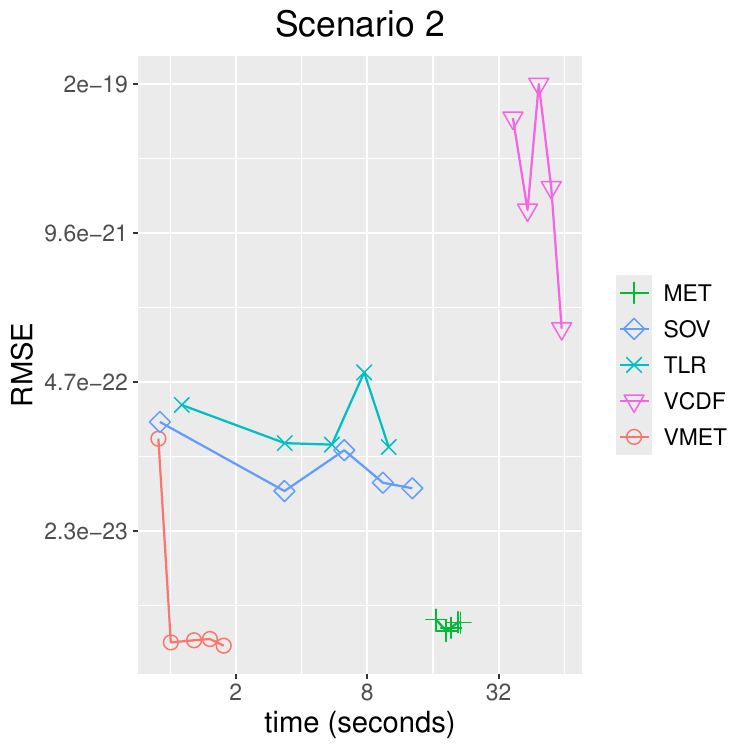}
    \end{subfigure}%
    \begin{subfigure}{.3\textwidth}
    \centering
    \includegraphics[width =.99\linewidth, trim=0 0 1in 0, clip]{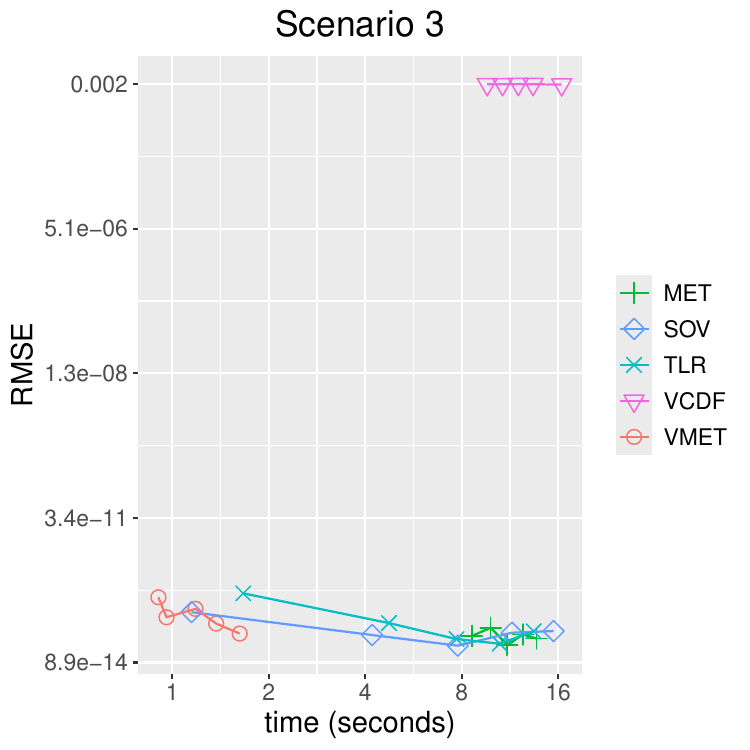}
    \end{subfigure}%
    \begin{subfigure}{.08\textwidth}
    \centering
    \includegraphics[width =.99\linewidth, trim=4in 0 0 0, clip]{plots/err_vs_time_lowdim_3_3.pdf}
    \end{subfigure}%
    \caption{For the scenarios in Table~\ref{tbl:low_dim_exp} with $n=900$, we show the RMSE of 30 probability estimates against the computation time (both on a log scale), for VMET, VCDF, TLR, SOV, and MET. VMET is run with $m \in \{10, 30, 50, 70, 90\}$ and $N = 10^{4}$, SOV and TLR are run with $N \in \{1, 3, 5, 7, 9\} \times 10^{4}$, and MET and VCDF are run with $N \in \{6, 7, 8, 9, 10\} \times 10^{3}$. An MET estimate with $N = 10^{7}$ is used as the truth.}
    \label{fig:low_dim_exp}
\end{figure*}

We compared the methods listed at the beginning of Section~\ref{sec:numerical} on the same three scenarios in Table~\ref{tbl:low_dim_exp} used in Section~\ref{subsec:var_reorder} with $n=900$. The resulting RMSE and average computation time of thirty probability estimates for the three MVN problems are plotted in Figure~\ref{fig:low_dim_exp}. 
The log scale was needed to accommodate the substantially different accuracy levels. The true values of the three MVN probabilities are unknown, but the MET method can be considered the gold standard. (But note that MET is not computationally scalable to large $n$.) {For our VMET method, we gradually increased $m$ to demonstrate the trade-off between estimation bias and computation efficiency. The implementations of VMET, SOV, and TLR store MC samples of $h$, whose average becomes the MVN probability estimate, in a mantissa-exponent format, hence avoiding numerical underflow}. 

Overall, VMET had the best trade-off between computation cost and accuracy, achieving the same level of errors as MET while reducing the computation times by 90\%. VCDF had the largest estimation errors, followed by TLR and SOV. TLR under-performed SOV because the TLR Cholesky factor of $\bfSigma$ introduced bias and the block variable reordering used in TLR is less effective in variance reduction than the univariate reordering. The third example had centered integration limits, where the optimal $\hat{\bfgamma}$ for the proposal density used in MET and VMET was simply zero, hence the MET integrands were identical to the SOV integrands. { For MET, solving for $\hat{\bfgamma}$ dominated the computation cost, as seen from the computation time difference between MET and SOV, both at $N = 10{,}000$, although the times were based on different implementations. Utilizing our proposed linear-complexity gradient and Hessian estimators in \eqref{equ:grad} and \eqref{equ:hessian}, the optimization cost of $\bfgamma$ was significantly reduced, enhancing the scalability of the importance-sampling technique proposed in \cite{botev2017normal}.} We show in the next section that due to our complexity advantage, which is $\order(n)$ and $\order(n^{0.5})$ lower than SOV and TLR, respectively, our computation times were lower than those of SOV and TLR when $n$ was larger. The estimation bias of our method can be significant when $m$ is chosen too small (e.g., $m = 10$ in Figure \ref{fig:low_dim_exp}), but a reasonably large $m$ (e.g., between $30$ and $50$) was sufficient to achieve negligible bias in our comparisons, even in tens of thousands of dimensions.

\subsection{Comparison for high-dimensional MVN probabilities}
\label{subsec:high_dim_numerical}

\begin{figure*}[h!]
    \centering
    \begin{subfigure}{.3\textwidth}
    \centering
    \includegraphics[width =.99\linewidth, trim=0 0 1in 0, clip]{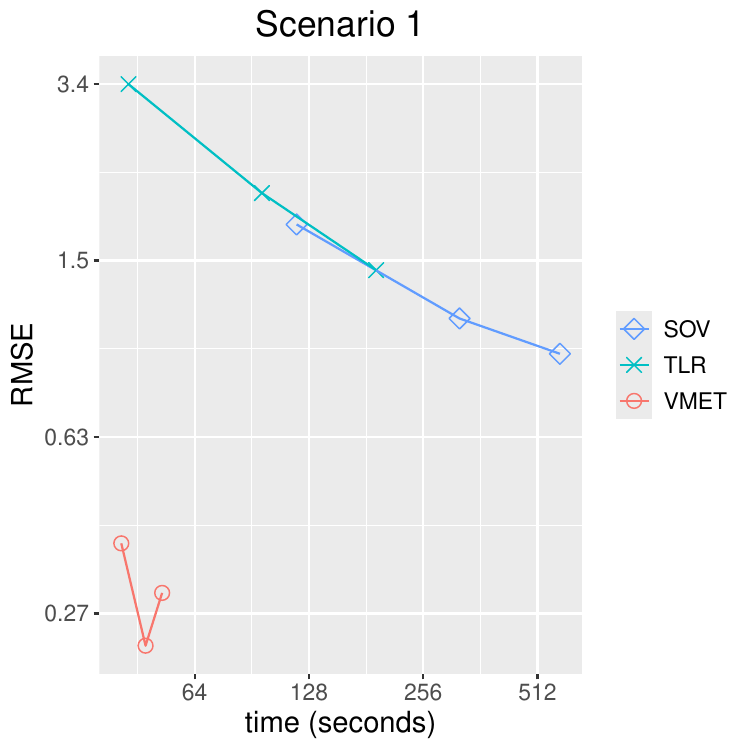}
    \end{subfigure}%
    \begin{subfigure}{.3\textwidth}
    \centering
    \includegraphics[width =.99\linewidth, trim=0 0 1in 0, clip]{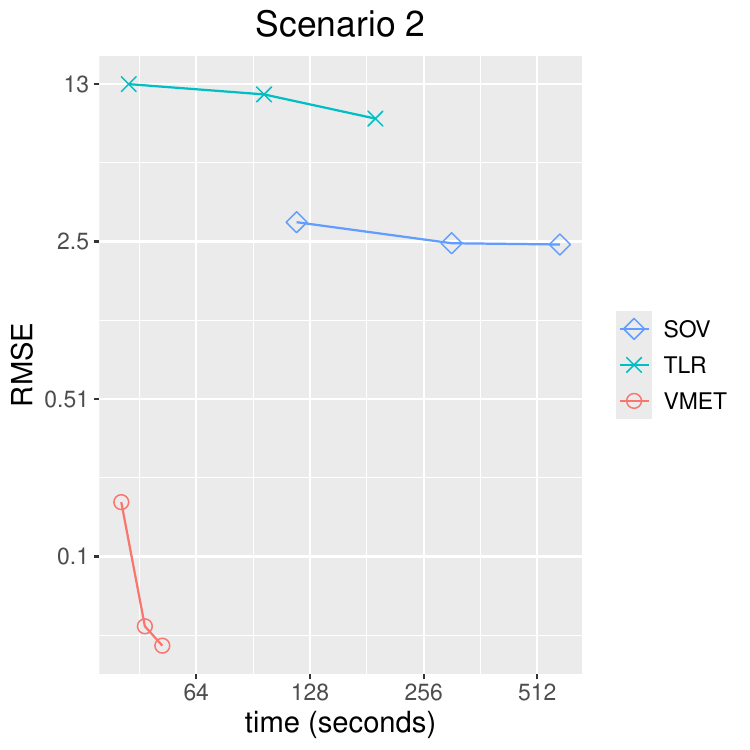}
    \end{subfigure}%
    \begin{subfigure}{.3\textwidth}
    \centering
    \includegraphics[width =.99\linewidth, trim=0 0 1in 0, clip]{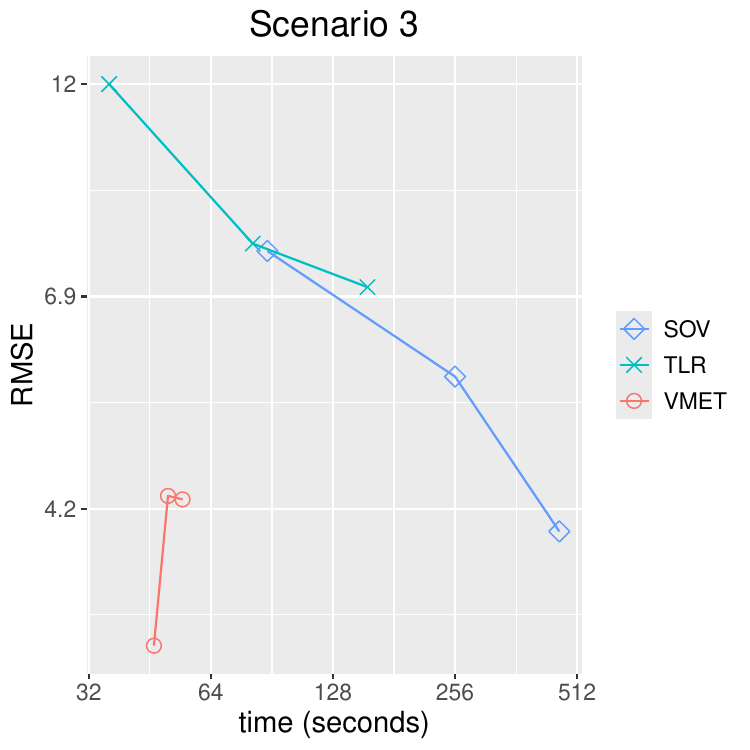}
    \end{subfigure}%
    \begin{subfigure}{.08\textwidth}
    \centering
    \includegraphics[width =.99\linewidth, trim=4in 0 0 0, clip]{plots/err_vs_time_highdim_exp1_new.pdf}
    \end{subfigure}%
    
    \caption{For the scenarios in Table~\ref{tbl:low_dim_exp} with $n=6{,}400$, we show the RMSE of 30 log-probability estimates against the computation time (both on a log scale), for VMET, TLR, and SOV. VMET is run with $m \in \{30, 50, 70\}$ and $N = 10^{5}$, SOV and TLR are run with $N \in \{2, 5, 10\} \times 10^{4}$. For Scenarios 1 and 2, the average of VMET estimates at $m = 70$ is used as truth. For Scenario 3, the average of SOV estimates at $N = 10^{5}$ is used as truth.}
    \label{fig:high_dim_exp}
\end{figure*}

For larger $n$, we only compared our method to TLR and SOV, as MET and VCDF are computationally infeasible. We considered three $n = 6{,}400$-dimensional MVN probabilities according to Scenarios~1 to 3 in Table~\ref{tbl:low_dim_exp}, respectively. For this high-dimensional study, the nugget parameter was increased from $0.01$ to $0.03$ to avoid singularity, while the other covariance parameters remained unchanged. The RMSE of log-probability estimates and computation times are shown in Figure~\ref{fig:high_dim_exp}. 

\looseness=-1
For Scenarios 1 and 2, we considered VMET with $m = 70$ as the benchmark, as VMET arguably has lower MC variance than SOV and TLR for tail MVN probabilities, and the bias caused by the Vecchia approximation was likely insignificant compared with the MC variance. For Scenario 3, we used SOV as benchmark, because for centered MVN probabilities, MET and SOV amount to the same integrand, hence SOV is expected to have the same accuracy as MET. { Compared to SOV and TLR, VMET significantly reduced the MC variance, hence improving estimation accuracy for tail MVN probabilities; for centered MVN probabilities, VMET also achieved a more favorable trade-off between estimation error and computation cost. Figure~\ref{fig:high_dim_exp_boxplot} in the Appendix provides more detailed information on the three methods' probability estimates.} Furthermore, TLR, the state-of-the-art in scalability, may suffer from unpredictable numerical singularity, whereas VMET is numerically stable, because unlike low-rank matrix approximations, the Vecchia approximation has a theoretical guarantee for positive definiteness {(see Prop.~1 in \citealp{Katzfuss2017a})}.

\subsection{{ Stress test for VMET under constant correlation}}
\label{subsec:const_corr_sim}

\looseness=-1
{ Constant correlation structure is arguably most challenging for the Vecchia approximation. Under constant correlation, each variable is equally correlated with another, which can result in relatively high approximation errors that are independent from the construction of the conditioning sets $\{c(i)\}_{i = 1}^{n}$ when $m$ is fixed. Therefore, the difference between the proposed VMET method and the MET method became more distinct here in Figure~\ref{fig:const_corr_exp} than under commonly used spatial covariance structures in Figure~\ref{fig:low_dim_exp}.
\begin{figure*}[h!]
    \centering
    \begin{subfigure}{.4\textwidth}
    \centering
    \includegraphics[width =.99\linewidth]{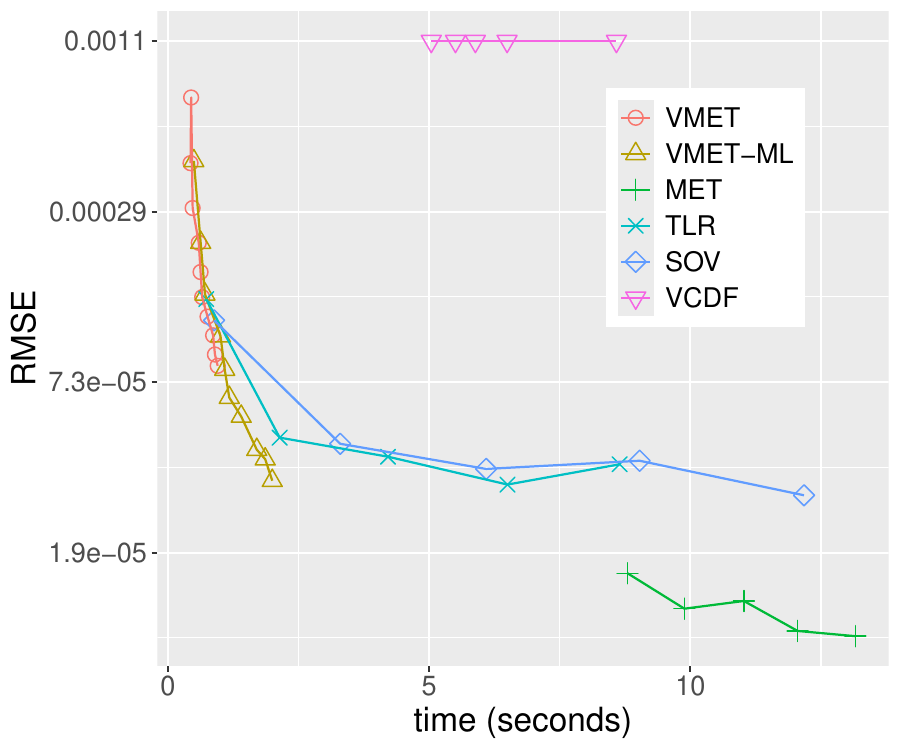}
    \end{subfigure}%
    \begin{subfigure}{.4\textwidth}
    \centering
    \includegraphics[width =.99\linewidth]{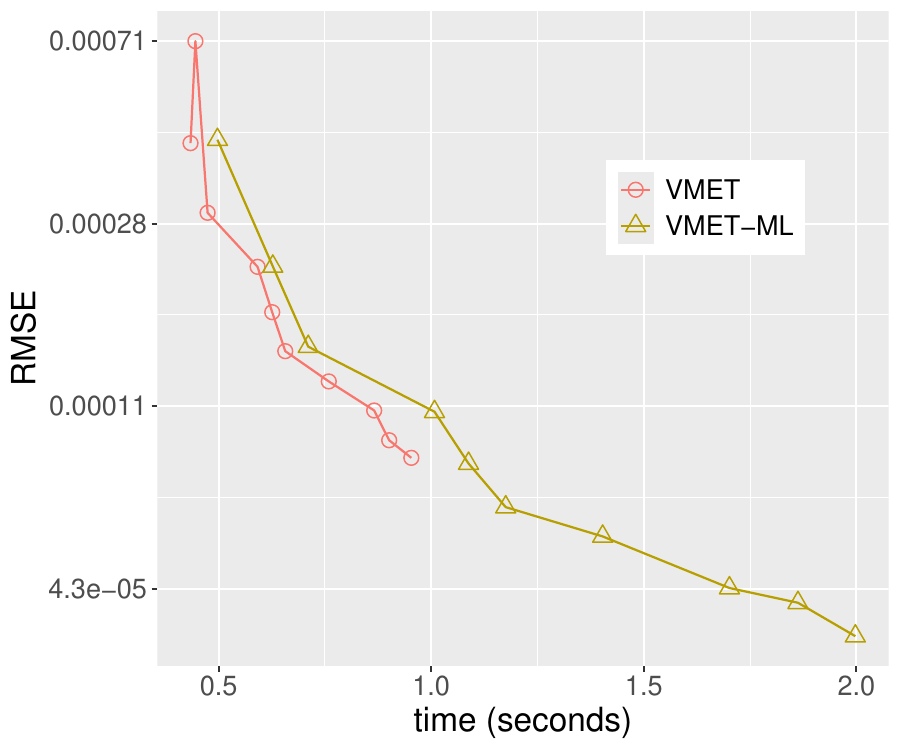}
    \end{subfigure}%
    
    \caption{The RMSE of 30 probability estimates (on a log scale) against computation time under a constant correlation structure. In this simulation, $n = 900$, $\ba = -\infty$, $\bb = \bfzero$, and $\bfSigma$ is a correlation matrix with a constant correlation of $0.5$. The left panel compares VMET, VMET with multi-level MC (VMET-ML), MET, TLR, SOV, and VCDF, while the right panel compares only VMET and VMET-ML to highlight their difference in the error-and-computation-cost trade-off. VMET is run with $m \in [10, 90]$ and $N = 10^{4}$, SOV and TLR are run with $N \in \{1, 3, 5, 7, 9\} \times 10^{4}$, and MET and VCDF are run with $N \in \{6, 7, 8, 9, 10\} \times 10^{3}$. The truth in this case is analytically available: $\Phi_n(\ba, \bb; \bfSigma) = 1/(n+1)$.}
    \label{fig:const_corr_exp}
\end{figure*}
Even under a most challenging covariance structure for the Vecchia approximation, VMET still provided a better trade-off between estimation errors and computation cost than all methods except for MET, which achieved higher accuracy with substantially more computation cost. VMET with multi-level MC used $m_{2}/2 = m_{1} = m$ and $10N_{2} = N_{1} = N$, which, despite higher accuracy than VMET at the same $m$, led to a less favorable trade-off between accuracy and computation cost. It is worth mentioning that the constant correlation structure is rarely used in practice while those in Table~\ref{tbl:low_dim_exp} are more representative, especially among spatial applications. Therefore, Figure~\ref{fig:const_corr_exp} serves as a `stress test' for the proposed VMET method, providing numerical support for its performance under general non-spatial covariance structures, while Figures~\ref{fig:low_dim_exp} and \ref{fig:high_dim_exp} are more indicative of its performance in various applications.
}

\subsection{Sampling from TMVN distributions}
\label{subsec:samp_TMVN}

\begin{figure*}[h!]
\centering
\hspace*{\fill}
\begin{subfigure}{.33\textwidth}
\centering
\includegraphics[width =.99\linewidth]{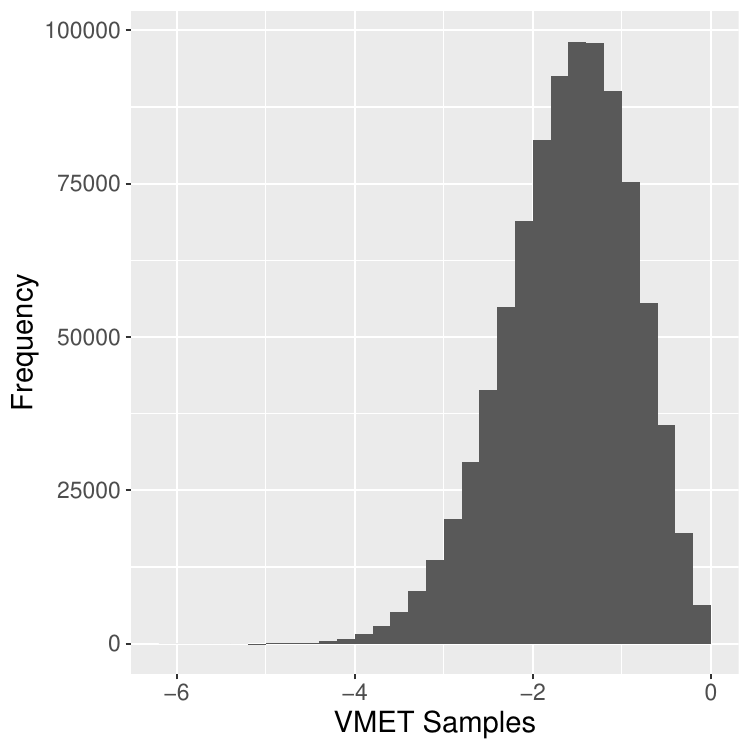}
\end{subfigure}%
\hfill
\begin{subfigure}{.33\textwidth}
\centering
\includegraphics[width =.99\linewidth]{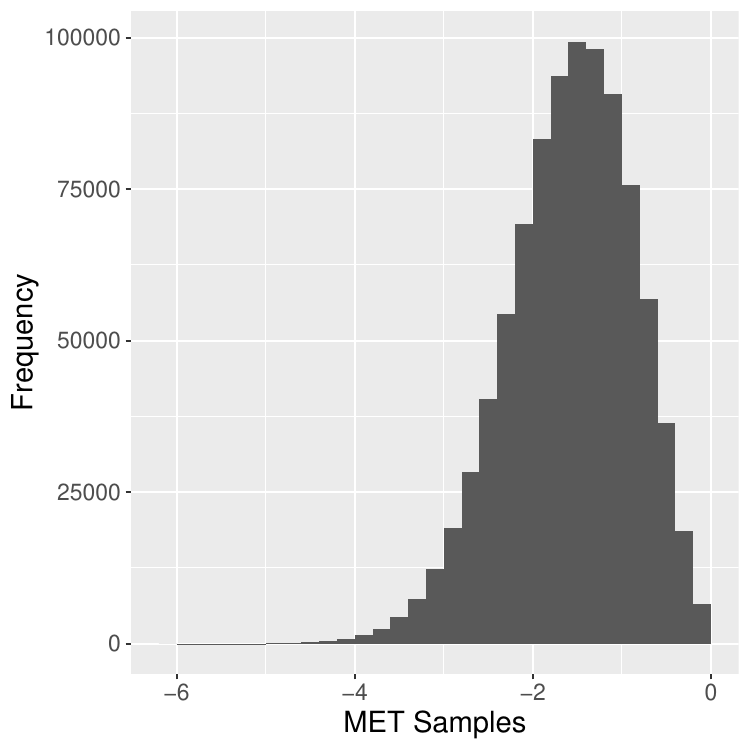}
\end{subfigure}%
\hfill
\begin{subfigure}{.33\textwidth}
\centering
\includegraphics[width =.99\linewidth]{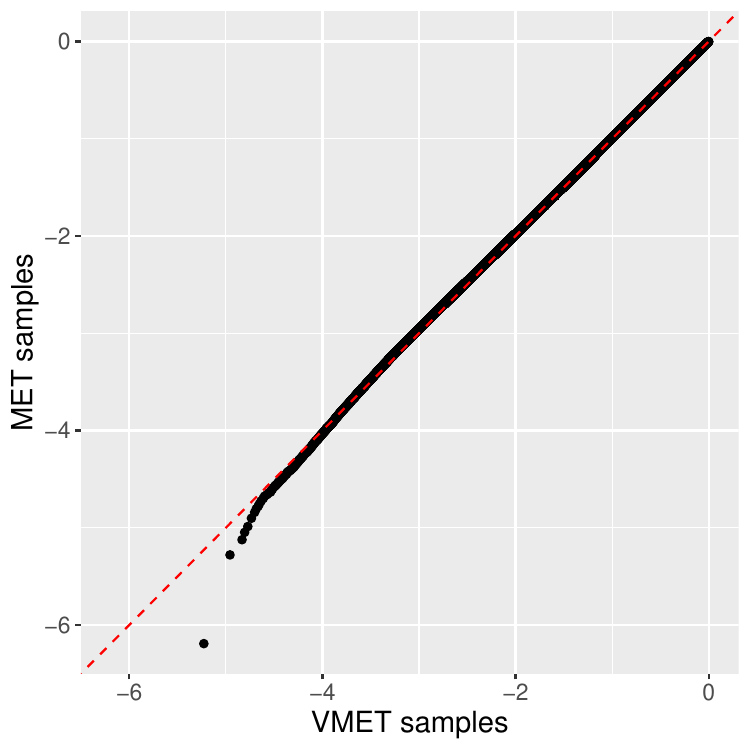}
\end{subfigure}%
\hspace*{\fill}
\caption{Histograms and q-q plot of the TMVN samples generated by our proposed VMET method and the MET method \citep{botev2017normal}}
\label{fig:TMVN_sample}
\end{figure*}

Our proposed VMET method can also be used for sampling TMVN distributions. We compared the performance of VMET with that of MET in terms of the quantile-quantile (q-q) plot of the drawn samples and the time used. For the TMVN distribution, we considered Scenario~1 in Table~\ref{tbl:low_dim_exp} with $n = 900$ and the same Mat\'ern kernel as in Sections~\ref{subsec:var_reorder} and \ref{subsec:low_dim_numerical}. We generated $N = 1{,}000$ samples using the VMET method with $m = 30$ described in Algorithm~\ref{alg:exp_tilt_vecchia} and the MET method implemented in the \texttt{TruncatedNormal} R package; a visual comparison is provided in Figure~\ref{fig:TMVN_sample}.

The two groups of samples aligned very well with each other except for the very tail region, because VMET essentially truncates the conditioning set compared with MET, which, in this example, amounted to conditioning on fewer left-truncated variables, resulting in slightly larger conditional means. In terms of computation time, MET used 494.4 seconds, whereas VMET only used 19.9 seconds. The time difference will become even more pronounced for larger $n$.

Note that the acceptance rate can be too low to draw exact samples from the TMVN distribution for moderately large $n$ (e.g., $n=10^3$), even with the exponential tilting proposal density \eqref{equ:exp_tilt_proposal}. Hence, for large $n$, we propose a regional sampling technique in Section~\ref{sec:application} for drawing samples only over a region of interest.

\section{Analysis of Partially Censored Data \label{sec:application}}

Censoring is a common issue in many application areas including environmental science, where sensors in earth, water, or air often have a threshold below which the quantity of interest cannot be detected. Assuming that the quantity of interest follows a joint normal distribution, the observed and censored data collected by the sensor jointly follow a censored MVN distribution. Mathematically, assuming a zero mean, the data $\bz = [z_1, \ldots, z_n]^{\top}$ follow a censored MVN distribution if
\[
\bx \sim \normal(\bfzero, \bfSigma), \qquad \text{ and } 
    z_{i} = x_{i}\ \mbox{if } x_{i} > b_{i} \mbox{ and }
    \mbox{NA}\ \mbox{otherwise},
\]
where $\bx$ is the quantity of interest that the sensors are attempting to measure, and $\{b_{i}\}$ are detection thresholds of the sensors. Without loss of generality, we assume that only the first $n_1$ locations are observed (i.e., $x_{i} > b_{i}$ if and only if $i \le n_1$).

\looseness=-1
Given the data $\bz$, the likelihood of the censored MVN distribution is:
\[
    f(\bz) = f_{\bx_{1:n_{1}}}(\bz_{1 : n_{1}}) \Pr(\bx_{n_{1} + 1 : n} < \bb_{n_{1} + 1 : n} \mid \bx_{1:n_{1}} = \bz_{1 : n_{1}}),
\]
{based on which maximum likelihood estimation can be used to estimate parameters in $\bfSigma$ (e.g., Mat\'ern covariance parameters)}. Under the Vecchia approximation, the above likelihood can be written as
\begin{align}
    f(\bz) = \Big( \prod_{i=1}^{n_{1}} f(x_{i} | \bx_{c(i)}) \Big) 
    \Big( \int_{-\bm{\infty}}^{\bb_{n_{1} + 1:n}} \prod_{i=n_{1} + 1}^n f(x_i | \bx_{c(i)}) \, \diff \bx_{n_{1} + 1:n} \Big). \label{equ:censor_mvn_dens}
\end{align}
\looseness=-1
Notice that $\bx_{c(i)}$ for $i > n_{1}$ may include observed $x_{j}$ (i.e., with $j \le n_{1}$). We can view \eqref{equ:censor_mvn_dens} as a limit case of \eqref{equ:MVN_prob_def} with $\ba_{1:n_1} \rightarrow \bz_{1:n_1} \leftarrow \bb_{1:n_1}$ and $\ba_{n_1+1:n} = -\bm{\infty}$.

\subsection{Parameter estimation}
\label{subsec:parm_est}

It is common practice in many environmental applications to substitute censored data with a constant value, such as half the level of detection (LOD), the LOD divided by the square root of 2, or simply zero \citep{croghan2003methods}. These LOD-based methods are also broadly used as a benchmark for more sophisticated Markov-chain-Monte-Carlo-based methods \citep{de2005bayesian, ordonez2018geostatistical}, which are often not scalable to large $n$. We show that in the presence of censored data, the censored MVN model in \eqref{equ:censor_mvn_dens} can achieve more accurate parameter estimation than LOD-based methods. We generated a spatial field $\bx$ over $6{,}400$ locations on a $80 \times 80$ grid in the unit square based on a GP with mean zero and a Mat\'ern covariance with variance, range, smoothness, and nugget of $(1.0, 0.1, 1.5, 0.03)$ in the same setup as Scenario 1 in Table~\ref{tbl:low_dim_exp}.
All $x_i$ values below $b_i = 0$ were censored to obtain the data $\bz$. Because the censoring thresholds were zero, all LOD-based methods discussed in \citep{croghan2003methods} amounted to the same data augmentation.
\begin{wrapfigure}{r}{0.5\textwidth}
\begin{center}
\includegraphics[width=0.49\textwidth]{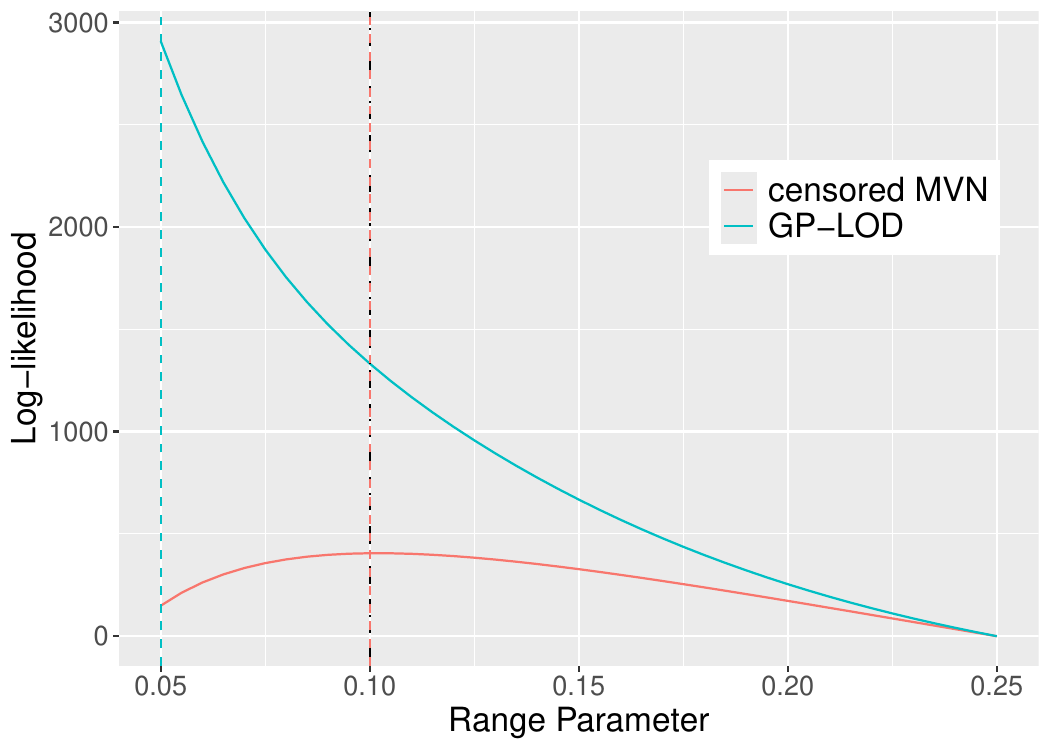}
\end{center}
\caption{Log-likelihood curves as a function of the range parameter for our censored MVN approach and for the LOD-augmented GP. Both curves were vertically shifted to have their minimums at zero.}
\label{fig:censored_normal_sim}
\end{wrapfigure}
We considered a specific realization, for which $2{,}585$ responses were censored and $3{,}815$ were observed. To visualize the log-likelihood manifolds of the censored MVN model and the LOD approach, we fixed variance, smoothness, and nugget to their true values while plotting the log-likelihood against the range parameter. The GP log-likelihood using LOD-augmented data peaked at the lower bound of the interval we considered, whereas the censored MVN yielded a log-likelihood curve that peaked at exactly the true value $0.1$, indicating the efficacy gain from properly including the information at censored locations. 
{For parameter estimation, we did not apply the variable reordering discussed in Section~\ref{sec:var_reorder} to avoid discontinuity in the log-likelihood surface, because both the proposal density $q(\bx; \bfgamma)$ and the likelihood ratio $h(\bx; \bfgamma)$ are computed w.r.t.\ a given ordering of the integration variables.}

\subsection{Sampling TMVN over regions of interest}
\label{subsec:sample_TMVN_region}

Once parameters have been estimated, we can sample the censored data given the observed data as in Section~\ref{subsec:samp_TMVN}, holding the first $n_1$ observed data fixed. Unlike for parameter estimation, we recommend using variable reordering for sampling TMVN distributions to improve the acceptance rate. However, the acceptance rate for sampling the TMVN distribution can be very low even with exponential tilting when the number of censored responses is high (e.g., $(n-n_1) > 1{,}000$), making it infeasible to obtain a reasonable number of samples for posterior inference. In this case, we propose to divide the censored locations according to subregions of the input space, each containing a manageable number (say, at most $1{,}000$) of censored locations; then we can draw TMVN samples over each region of interest separately. {The responses at censored locations over one subregion are assumed to be conditionally independent from those over another subregion, given all observed responses}. This idea is intuitive and simple to implement. {Its downside is that it ignores censored locations outside of the selected region. As a heuristic remedy, we extend the boundary of the region of interest when drawing TMVN samples to incorporate surrounding information; this extension reduces the bias but also the acceptance rate}. In many applications, obtaining the joint posterior distribution of the quantity of interest over a subregion is sufficient; for example, one could obtain the joint distribution of a water-related quantity over a catchment area and then compute the distribution of functions of interests (e.g., the average) as necessary.

\begin{figure*}[h!]
\centering
\hspace*{\fill}
\begin{subfigure}{.33\textwidth}
\centering
\includegraphics[width =.99\linewidth]{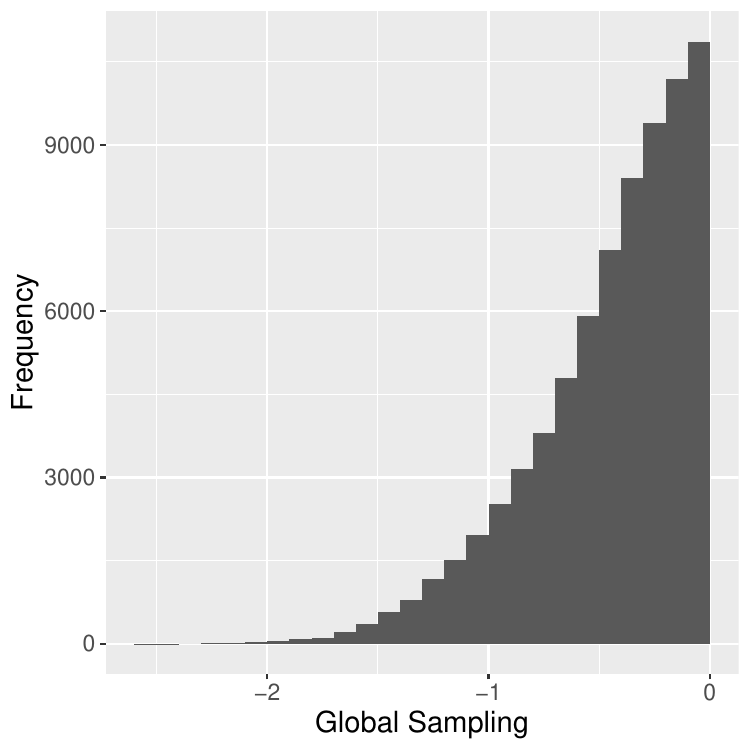}
\end{subfigure}%
\hfill
\begin{subfigure}{.33\textwidth}
\centering
\includegraphics[width =.99\linewidth]{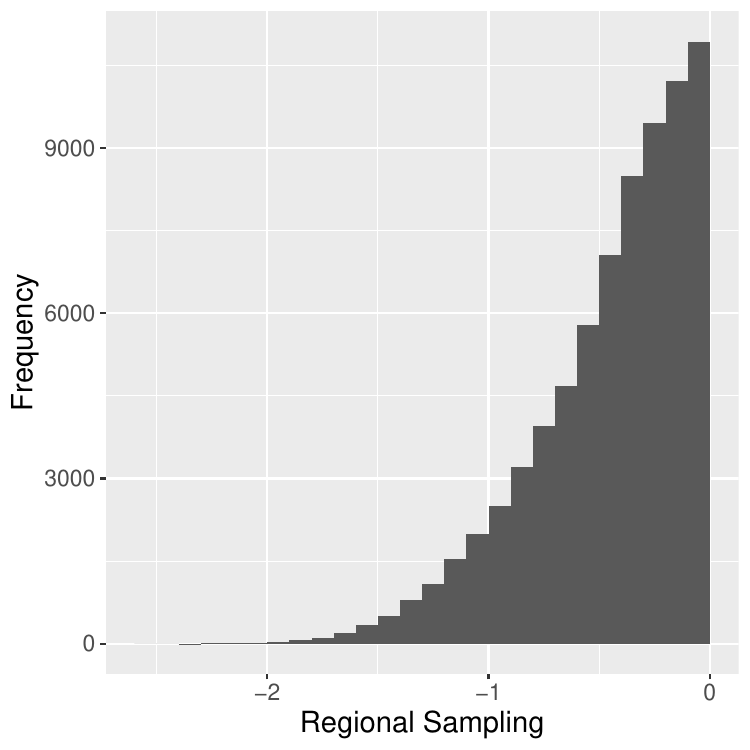}
\end{subfigure}%
\hfill
\begin{subfigure}{.33\textwidth}
\centering
\includegraphics[width =.99\linewidth]{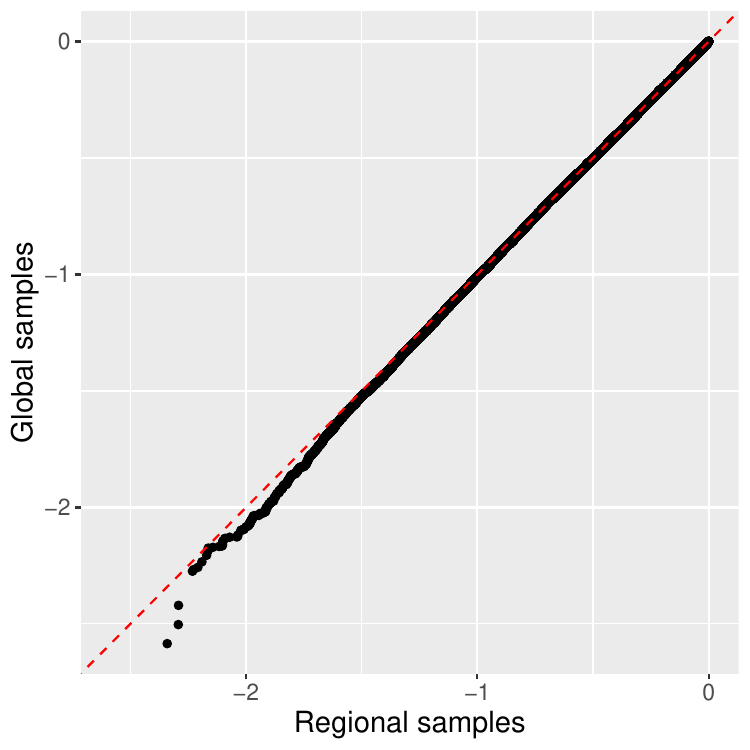}
\end{subfigure}%
\hspace*{\fill}
\caption{Histograms and q-q plot of the TMVN samples drawn by the `global' and `regional' schemes}
\label{fig:PTMVN_sample_low}
\end{figure*}

\looseness=-1
To demonstrate this idea, we first considered an example in sufficiently low dimensions so that the `global' simulation of all censored locations had a reasonable acceptance rate above $10^{-4}$. Specifically, we considered Scenario 1 in Table~\ref{tbl:low_dim_exp}, where the censoring threshold is zero. Figure~\ref{fig:PTMVN_sample_low} compares $N = 1{,}000$ samples of the north-west quarter (i.e., $[0, 0.5] \times [0.5, 1.0]$) generated by the global and regional sampling schemes using VMET. 
For the regional sampling scheme, we sampled the TMVN distribution defined over the region $[0, 0.6] \times [0.4, 1.0]$, which enveloped the region of our interest, conditional on all observed locations. For the global sampling scheme, we needed to sample from a $380$-dimensional TMVN distribution, whereas for the regional scheme, we only needed to sample from a $125$-dimensional TMVN distribution, resulting in a $50$ to $100$ times difference in acceptance rates. The two groups of samples were very similar except for minor differences in the tails, which can be attributed to the reduced number of censored locations used in the regional sampling scheme. Conditioning on more censored locations would decrease the conditional mean under our Mat\'ern covariance kernel. We also compared the RMSE for predicting the censored data in the north-west quarter, where the global and regional sampling produced errors of $0.24$ and $0.25$, respectively. In contrast, the GP with the LOD-augmented data produced a prediction error of $0.62$ and the stochastic approximation of expectation maximization (SAEM), a state-of-the-art method for sampling TMVN distributions from \cite{ordonez2018geostatistical}, produced an error of $0.47$. For SAEM, we set the MC sample size to $500$ as in \cite{ordonez2018geostatistical}, resulting in a computation time of approximately two hours, which was significantly higher than for our proposed methods. The regional scheme had much higher accuracy than the competing methods and the minor accuracy loss compared with the global scheme, we believe, is a good trade-off for the gain in computational efficiency.

Next, we considered an example in higher dimensions to demonstrate that the VMET method can be used in combination with the regional sampling scheme to improve posterior inference when there are censored GP responses. We again considered Scenario 1 from Table~\ref{tbl:low_dim_exp}, but now with $n = 6{,}400$. The region of interest and the region over which we drew TMVN samples were the same as in the low-dimensional example above. We had a total of $3{,}278$ censored locations, with $1{,}075$ inside our sampling region. $N = 1{,}000$ samples were drawn and used in combination with all observed responses to predict the responses at $500$ randomly selected testing locations in our region of interest (i.e., $[0, 0.5] \times [0.5, 1.0]$). For comparison, we predicted with the GP using the LOD-augmented data; we also attempted a comparison to SAEM, but the approach was computationally infeasible for sampling the $1{,}075$-dimensional TMVN distribution due to extremely low acceptance rates. The LOD-augmented GP produced an RMSE of $0.45$, whereas the RMSE of our method was only $0.19$. While the global approach cannot be computed as a gold-standard comparison method in this high-dimensional example, note that the nugget variance is $0.03$, indicating a minimum prediction RMSE of $0.17$. This highlights the benefits of properly taking the censored locations into consideration for posterior inference. In this example, the acceptance rate was approximately $2 \times 10^{-5}$ and the sampling took less than 44 minutes. Combined with the results from Section~\ref{subsec:parm_est}, the VMET algorithm outlined in Algorithm~\ref{alg:exp_tilt_vecchia} enables scalable (linear-complexity) computation for the partially censored MVN model in terms of parameter estimation and posterior inference, significantly outperforming the LOD-augmented GP and the SAEM approaches.

\subsection{Analysis of groundwater contamination}

We considered groundwater tetrachloroethylene concentrations from the United States Geological Survey (USGS) to demonstrate that our proposed VMET method can be used in combination with the censored MVN model to enhance the modeling of partially censored data. Tetrachloroethylene can be used in dry cleaning, manufacturing other chemicals, and cleaning metals, but exposure to tetrachloroethylene is harmful and may cause skin and respiratory problems. The data were collected at irregular locations across the United States over the time period 2000 to 2022. In this section, the locations $\{\bs_{i} \in \mathbb{R}^3\}$ are spatial-temporal coordinates consisting of longitudes, latitudes, and dates. Overall, there are $n = 24{,}701$ spatio-temporal responses, among which there are $20{,}730$ censored responses due to detection limits. The detection thresholds are known and vary across spatio-temporal locations. Environmental contaminants are usually modeled using MVN distributions after a logarithm transformation \citep{helsel1990less}. During data preprocessing, we first applied a log-transform to both observations and detection thresholds. Next, we normalized the transformed observations and the detection thresholds with the mean and standard deviation of the transformed observations. The spatial and temporal coordinates were linearly scaled into the unit hypercube in $\mathbb{R}^{3}$, with the two spatial coordinates scaled by the same constant. {The dataset, including observations and detection thresholds, is randomly split into 80\% training and 20\% testing.}

The responses and detection thresholds after transformation were modeled by a partially censored MVN distribution with zero mean and a Mat\'ern-1.5 kernel. We used one range parameter for longitude and latitude and another one for the temporal coordinate. Four parameters $(\sigma^2, \beta_{1}, \beta_{2}, \tau^2)$, namely variance, spatial range, temporal range, and nugget, were estimated by maximizing the likelihood \eqref{equ:censor_mvn_dens} as described in Section~\ref{subsec:parm_est}. Notice that we used the chordal distance for our study since the spatial locations are all within the United States, where the chordal distance closely resembles the great-circle distance. For optimization, we set $m = 50$, MC sample size to $10^4$, and used the robust Nelder and Mead Algorithm \citep{nelder1965simplex} native to R, which took approximately five hours to reach convergence; the parameters were estimated to be $(13.28, 0.04, 12.46, 0.07)$, indicating very strong temporal correlation. 

\begin{figure*}[h!]
    \centering
    \begin{subfigure}{.99\textwidth}
    \centering
    \includegraphics[trim = 0 0 7in 0, clip, height = 2in]{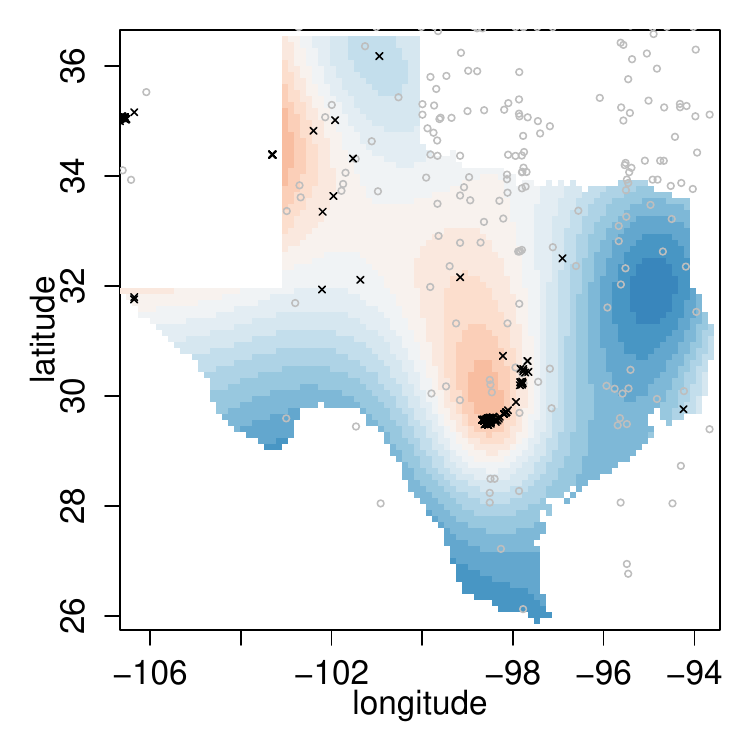}
    \includegraphics[trim = 6in 0 0 0, clip, height = 2in]{plots/PCE_modeling.pdf}
    \end{subfigure}%
    \caption{Inference (i.e., posterior means) on the logarithm of the groundwater tetrachloroethylene concentrations in Texas made by (left) partially censored MVN model and (right) LOD-augmented GP. Black crosses and grey circles denote the spatial locations of observed and censored data, respectively.}
    \label{fig:PCE_modeling}
\end{figure*}

We used the state of Texas as our region of interest for posterior inference. Specifically, our goal was to make predictions over a dense regular $0.13^\circ \times 0.11^\circ$  longitude-latitude grid covering Texas on the last day of the dataset's temporal range. Based on the predictions, we produced a `heatmap' for the groundwater tetrachloroethylene concentrations in Texas. Conditional on the training observations, we drew $N = 1{,}000$ samples of the TMVN distribution defined over the $546$ censored spatio-temporal points in the training dataset that are located in Texas. Inference over the grid as well as the testing spatio-temporal points in Texas was made by kriging using observed responses and the TMVN samples drawn previously. For comparison, we also considered predictions made by a GP using the augmented data where censored responses were replaced by their LODs. The SAEM method was computationally infeasible for sampling from the $546$-dimensional TMVN distribution, and it was hence excluded from our comparison. Figure~\ref{fig:PCE_modeling} compares the heatmaps produced by the partially censored MVN model and the GP model.
The heatmap generated by the partially censored MVN model exhibited larger variations across space and more details, while the inference produced by the GP with augmented data appeared unnatural, with predictions almost constant over large regions; hence, this approach may not be suitable for the logarithm of the contaminant levels \citep{helsel1990less}. The LOD-augmented GP is likely to have significantly overestimated the responses over the grid, which may have meaningful practical importance, as the two approaches differ in their assessment of where tetrachloroethylene concentrations are elevated and hence where mitigation may be necessary. {To compare prediction accuracy at the $179$ spatio-temporal test points in Texas, we considered predictions of the binary indicators of whether responses where below or above the LOD, because the true concentration levels for the censored responses were unknown. In this comparison, the partially censored MVN model and the LOD-augmented GP had RMSEs (i.e., Brier scores) of 0.36 and 0.74, respectively. This indicates that the partially censored MVN model, powered by VMET, significantly improved inference accuracy for the tetrachloroethylene concentration dataset.}

\section{Conclusions \label{sec:conclusions}}

We proposed a linear-complexity method for estimating multivariate normal (MVN) probabilities and sampling truncated multivariate normal (TMVN) distributions at the same convergence and acceptance rate as the cubic-complexity state-of-the-art minimax exponential tilting (MET) method \citep{botev2017normal}. Specifically, our proposed VMET approach improves over the existing MET in three aspects. First, we re-parameterized the MET integrand, reducing the complexity of each Monte Carlo (MC) sample from $\order(n^2)$ to $\order(n)$, where $n$ is the dimension of the MVN probability. Second, we proposed a linear-complexity estimator for the gradient and Hessian of the logarithm of the density ratio based on the Vecchia approximation, which reduced the computation cost of finding the proposal density used in the MET method by $\order(n^2)$. Finally, we introduced a Vecchia-based variable reordering that achieves the same effectiveness as the classic univariate reordering \citep{gibson1994monte} used in the MET method for improving the convergence/acceptance rate, while reducing its complexity by a factor of $\order(n)$. 
We also compared our proposed VMET to the tile-low-rank (TLR) method from \cite{cao2021exploiting} that is currently the most scalable method for estimating MVN probabilities. In addition to higher estimation accuracy, VMET has lower time complexity (by a factor of $\order(n^{0.5}))$ and was faster (and more numerically stable) in our numerical examples. For sampling TMVN distributions, we proposed a `regional' sampling scheme to make posterior inference more scalable. To demonstrate the practical usage of the VMET method, we considered a partially censored GP approach that can jointly model observed and censored data. Explicitly utilizing censored data substantially improves the estimation and inference of the underlying GP compared to GP regression with data augmentation or Markov-chain-Monte-Carlo-based methods. {Since multivariate Student-$t$ (MVT) probabilities can be viewed as scale-mixture of MVN probabilities, the developed VMET method also allows estimating MVT probabilities and sampling truncated MVT (TMVT) distributions at the same complexity while maintaining the convergence/acceptance rate at the same level as MET}.

Similar to other scalable methods \citep[e.g.,][]{genton2018hierarchical, cao2021exploiting, nascimento2022vecchia}, our proposed VMET method amounts to an approximation to the covariance in the original MVN probability, and hence estimation bias is unavoidable. However, 
the Vecchia estimation bias is typically negligible relative to the MC errors for small values of the conditioning-set size $m$ (e.g., $30$ to $50$), making the bias from the Vecchia approximation a worthwhile trade-off for much higher computation scalability. 
Furthermore, the Vecchia approximation used in VMET provides parallel-computation capability with respect to the number of responses $n$, in addition to the parallelization with respect to the sample size $N$. One future research avenue is studying other families of proposal densities. Generally speaking, the design of proposal densities should consider how to 1) draw samples from the proposal density and 2) optimize the density function parameters. The convex-concave property, described in Proposition~\ref{prp:convex_concave}, for other families of proposal densities may no longer hold, necessitating innovative criteria for optimizing the proposal density, similar to \eqref{equ:gamma_hat}. {This paper only considers MVN and MVT probabilities with positive-definite covariance matrices; a direction for future research is to study Vecchia approximation of MVN probabilities subject to linear constraints, $\mbox{Pr}(\ba \le \bG \bx \le \bb)$, where $\bG$ can be rectangular, resulting in rank-deficient covariance matrices}.

Our \texttt{R} package \if1\blind{\texttt{VeccTMVN}}\fi \if0\blind {[blinded R package name]} \fi, which is available on CRAN, implements the methods proposed in this paper, including variable reordering, computing MVN/MVT probabilities, sampling TMVN/TMVT distributions, and multi-level MC. The code for reproducing the results in this paper using \texttt{R} package \if1\blind{\texttt{VeccTMVN} }\fi \if0\blind {[blinded R package name]} \fi can be found at\if1\blind
{
  \url{https://github.com/JCatwood/TMVN_Vecchia}.
} \fi 
\if0\blind
{
  Github.
} \fi

\footnotesize
\appendix
\section*{Acknowledgments}

The authors were partially supported by National Science Foundation (NSF) Grants DMS--1654083 and DMS--1953005. Support of JC's research was also provided by the University of Houston. Support for MK's research was also provided by the Office of the Vice Chancellor for Research and Graduate Education at the University of Wisconsin--Madison with funding from the Wisconsin Alumni Research Foundation. We would like to thank Kyle Messier for providing the tetrachloroethylene data and for several helpful comments and discussions.

\section*{Disclosure Statement}

The authors have no competing interests to declare.


\bibliographystyle{agsm}

\bibliography{mendeley,additionalrefs}
\end{document}